\documentclass[journal]{IEEEtran}
\usepackage{cite}
\usepackage{amsmath,amssymb,amsfonts,empheq}
\usepackage{algorithmic}
\usepackage{graphicx}
\usepackage{textcomp}
\usepackage{enumerate}
\usepackage{stfloats}
\usepackage{multirow}
\usepackage{booktabs}
\usepackage{algorithm}
\usepackage{color}
\usepackage{array}
\usepackage{makecell}
\usepackage{url}
\usepackage{booktabs}
\usepackage{etoolbox}
\usepackage{hyperref}
\usepackage{listings}
\usepackage{tikz}
\usepackage{microtype}
\usepackage{url}
\usepackage{tabularx}
\usepackage{subcaption}
\definecolor{SynColor}{RGB}{210, 220, 255} % Light Blue
\definecolor{SynBorder}{RGB}{100, 100, 200}
\definecolor{HILColor}{RGB}{210, 255, 210} % Light Green
\definecolor{HILBorder}{RGB}{100, 200, 100}
\definecolor{OTAColor}{RGB}{255, 210, 210} % Light Red
\definecolor{OTABorder}{RGB}{200, 100, 100}
\definecolor{OthColor}{RGB}{230, 230, 230} % Light Gray
\definecolor{OthBorder}{RGB}{150, 150, 150}
\definecolor{WorkColor}{RGB}{200, 240, 255} % Light Cyan
\definecolor{WorkBorder}{RGB}{50, 150, 200}

\definecolor{codegreen}{rgb}{0,0.6,0}
\definecolor{codegray}{rgb}{0.5,0.5,0.5}
\definecolor{codepurple}{rgb}{0.58,0,0.82}
\definecolor{backcolour}{rgb}{0.98,0.98,0.98}
\definecolor{stringblue}{rgb}{0.0, 0.0, 1.0} 
\definecolor{keycolor}{rgb}{0.54, 0.17, 0.89}

\lstdefinelanguage{json}{
    morekeywords={true,false,null},
    sensitive=true,
    morestring=[b]",
    morecomment=[l]{//},
    morecomment=[s]{/*}{*/},
    commentstyle=\color{codegreen},
    keywordstyle=\color{blue},
    stringstyle=\color{stringblue}
}

\lstdefinestyle{jsonstyle}{
    language=json,
    backgroundcolor=\color{backcolour},
    commentstyle=\color{codegreen},
    keywordstyle=\color{blue},
    numberstyle=\tiny\color{codegray},
    stringstyle=\color{stringblue},
    basicstyle=\ttfamily\footnotesize,
    breakatwhitespace=false,
    breaklines=true,
    captionpos=b,
    keepspaces=true,
    numbers=none, 
    numbersep=5pt,
    showspaces=false,
    showstringspaces=false,
    showtabs=false,
    tabsize=2,
    literate=
      *{:}{{{\color{keycolor}:}}}{1}
       {,}{{{\color{keycolor},}}}{1}
       {\{}{{{\color{keycolor}\{}}}{1}
       {\}}{{{\color{keycolor}\}}}}{1}
       {[}{{{\color{keycolor}[}}}{1}
       {]}{{{\color{keycolor}]}}}{1},
    escapeinside={\%*}{*)}
}

\ifCLASSINFOpdf
\else
\fi

\begin{document}

\title{CSRD2025: A Large-Scale Synthetic Radio Dataset for Spectrum Sensing in Wireless Communications}

\author{Shuo~Chang~\IEEEmembership{Member,~IEEE,} Rui Sun~\IEEEmembership{Student Member,~IEEE,}
  Jiashuo~He~\IEEEmembership{Member,~IEEE,} Sai Huang~\IEEEmembership{Member,~IEEE,} Kan Yu~\IEEEmembership{Member,~IEEE,} and Zhiyong~Feng~\IEEEmembership{Senior Member,~IEEE}
  \thanks{Shuo Chang (Corresponding Author), Jiashuo He, Sai Huang, Kan Yu, and Zhiyong Feng are the Key Laboratory of
    Universal Wireless Communications, Ministry of Education, Beijing University
    of Posts and Telecommunications, Beijing 100876, China,
    E-mail: \{changshuo, hejiashuo, huangsai, kanyu, fengzy\}@bupt.edu.cn.}
  \thanks{Rui Sun is the Shenzhen Future Network of Intelligence Institute and Guangdong Provincial Key Laboratory of Future Networks of
    Intelligence, The Chinese University of Hong Kong (Shenzhen), Shenzhen 518172, China,
    E-mail: ruisun@link.cuhk.edu.cn. Equal contributions mainly for supplying the MATLAB license.}
  \thanks{This work is supported in part by Youth Fund of the National Natural Science Foundation of China (62201090), in part by Basic Research Funds Project of Beijing University of Posts and Telecommunications under Grant (2024RC07).}}

\maketitle

\begin{abstract}
  The development of Large AI Models (LAMs) for wireless communications, particularly for complex tasks like spectrum sensing, is critically dependent on the availability of vast, diverse, and realistic datasets. Addressing this need, this paper introduces the ChangShuoRadioData (CSRD) framework, an open-source, modular simulation platform designed for generating large-scale synthetic radio frequency (RF) data. CSRD simulates the end-to-end transmission and reception process, incorporating an extensive range of modulation schemes (100 types, including analog, digital, OFDM, and OTFS), configurable channel models featuring both statistical fading and site-specific ray tracing using OpenStreetMap data, and detailed modeling of realistic RF front-end impairments for various antenna configurations (SISO/MISO/MIMO). Using this framework, we characterize CSRD2025, a substantial dataset benchmark comprising over 25,000,000 frames (approx. 200TB), which is approximately 10,000 times larger than the widely used RML2018 dataset. CSRD2025 offers unprecedented signal diversity and complexity, specifically engineered to bridge the Sim2Real gap. Furthermore, we provide processing pipelines to convert IQ data into spectrograms annotated in COCO format, facilitating object detection approaches for time-frequency signal analysis. The dataset specification includes standardized 8:1:1 training, validation, and test splits (via frame indices) to ensure reproducible research. The CSRD framework is released at \href{https://github.com/Singingkettle/ChangShuoRadioData}{\url{https://github.com/Singingkettle/ChangShuoRadioData}}\footnote{The CSRD simulation framework code is open-source. However, the specific 200TB CSRD2025 dataset instance characterized herein is not hosted for direct download due to the prohibitive costs associated with storage and bandwidth for data of this scale. The dataset is designed to be fully reproducible using the provided framework, configurations, and fixed random seeds available in the repository.} to accelerate the advancement of AI-driven spectrum sensing and management.
\end{abstract}

\begin{IEEEkeywords}
  Synthetic Radio Dataset, Spectrum Sensing, Large AI Models.
\end{IEEEkeywords}

\IEEEpeerreviewmaketitle

%======================================================================
% SECTION I: Introduction 
%======================================================================
\section{Introduction}
\label{sec:introduction}
\IEEEPARstart{T}{he} remarkable success of large-scale deep learning models, or Large AI Models (LAMs), particularly in natural language processing (NLP) and computer vision (CV)\cite{2024gpt4, 2025deepseekv3, 2024gemini1.5, 2020jared, 2022blip, 2022coca}, has spurred interest in their application to the complex domain of wireless communications\cite{2024bariah}. LAMs offer significant potential for tackling persistent challenges such as enhancing spectrum efficiency, optimizing resource allocation, and enabling more adaptive network operations by learning intricate patterns from vast amounts of data.

A fundamental challenge in modern wireless systems is efficient \textit{spectrum sensing}. As wireless devices proliferate and spectral congestion increases, the ability to accurately detect and characterize signals within the radio frequency (RF) spectrum becomes crucial for cognitive radio, dynamic spectrum access, and interference mitigation~\cite{2023chang, 2024xing}. While traditional sensing methods like energy detection~\cite{2016boulogeorgos} and cyclostationary feature detection~\cite{2015sepidband}, along with conventional machine learning~\cite{2013karaputugala}, have their merits, they often struggle with the dynamism and heterogeneity of contemporary wireless environments~\cite{2024guimaraes}.

LAMs, with their capacity to process extensive spectral data and learn complex spatio-temporal features, present a promising approach to overcome these limitations~\cite{2024bariah}. However, the success of LAMs is intrinsically tied to \textit{Scaling Laws}, which demonstrate that model performance scales predictably with increases in model size  ($N$), dataset size ($D$), and computational resources ($C$)~\cite{2020jared}. A key finding is the relationship $D \propto N^{0.74}$ to maintain optimal learning efficiency and prevent overfitting. This observation highlights a critical impediment to the broad application of LAMs in wireless communications: the acute need for massive, high-quality, and diverse datasets that accurately reflect the multifaceted complexities of real-world RF environments.

Obtaining such datasets in the wireless domain faces unique and substantial hurdles, diverging significantly from the relatively mature data acquisition pipelines in CV and NLP. As detailed in Section~\ref{sec:challenges_related_work}, collecting extensive real-world (Over-the-Air, OTA) spectrum data is hampered by the high cost of specialized RF equipment, the need for deep domain expertise, the non-stationary nature of the spectrum, and severe difficulties in obtaining accurate, fine-grained annotations~\cite{2023toolqa, 2025llmgenerated, Nguyen2024WRISTWidebandRealTime, Wood2022DeepLearningObject}. These challenges make generating the terabyte-scale, meticulously labeled datasets required by LAMs through purely OTA methods often impractical or prohibitively expensive.

\begin{figure*}[t]
  \centering
  \includegraphics[width=0.8\linewidth]{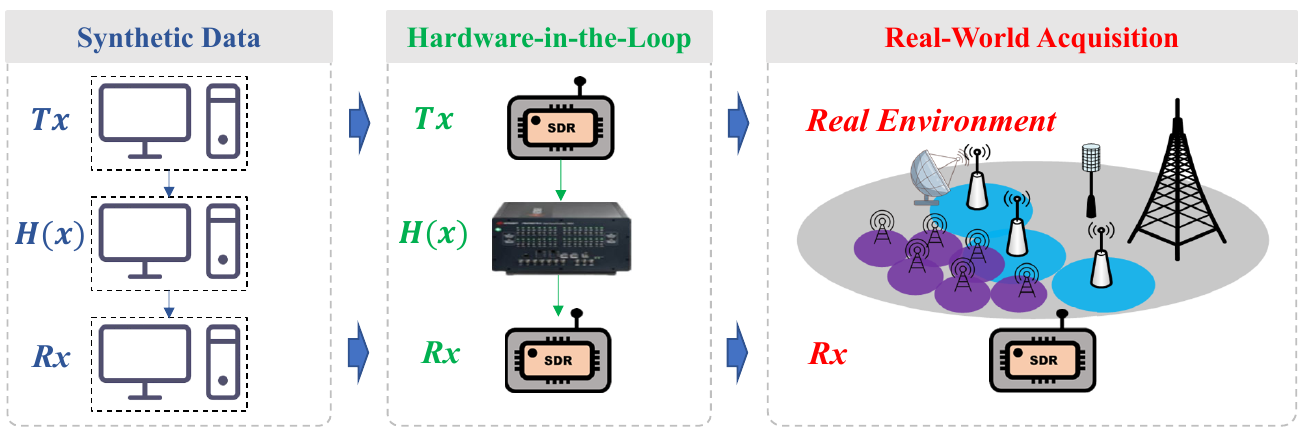}
  \caption{Proposed pragmatic three-stage strategy for constructing large-scale spectrum datasets. This approach balances feasibility and fidelity, starting with scalable \textbf{Synthetic Data Generation} (Stage 1, focus of this work), moving to \textbf{Hardware-in-the-Loop (HIL)} emulation (Stage 2), and potentially culminating in targeted \textbf{Real-World Acquisition} (Stage 3).}
  \label{fig:three_stage}
\end{figure*}

To pragmatically navigate these constraints, we advocate a multi-stage strategy for spectrum dataset construction, as illustrated in Fig.~\ref{fig:three_stage}. This approach commences with (1) scalable \textbf{Synthetic Data Generation}, which can then be complemented by (2) \textbf{Hardware-in-the-Loop (HIL)} emulation, and potentially culminate in (3) targeted \textbf{Real-World Acquisition} for final model refinement and validation. The present work focuses on establishing a robust and accessible foundation for the critical first stage: the generation of high-quality, large-scale synthetic spectrum data.

To this end, we introduce the ChangShuoRadioData (CSRD) framework, a novel, open-source, and modular simulation platform architected for this specific purpose. The CSRD framework emulates the end-to-end wireless transmission and reception chain, incorporating detailed physical layer modeling. Its design emphasizes three core pillars of innovation to enhance realism and utility for LAMs: \textbf{i) Comprehensive Signal and Impairment Modeling:} It supports an extensive library of approximately 100 distinct modulation schemes—spanning analog, diverse digital single-carrier formats (ASK, PSK, QAM, FSK, CPM), multi-carrier systems (OFDM, SCFDMA), and advanced waveforms like OTFS—and integrates detailed models of crucial RF front-end impairments (e.g., phase noise, amplifier nonlinearities, IQ imbalance) for both transmitters and receivers. \textbf{ii) Advanced Channel Emulation:} Beyond statistical channel models (e.g., Rayleigh, Rician), CSRD distinctively incorporates site-specific \textit{ray tracing} capabilities, leveraging OpenStreetMap (OSM) data to simulate propagation in realistic 3D environments. \textbf{iii) Scalability and Configurability:} The framework is designed for generating massive datasets with flexible control over scenario parameters, including multi-emitter configurations with controlled spectral overlap to study interference.

Leveraging the CSRD framework, we define and characterize CSRD2025, a large-scale synthetic dataset benchmark comprising over 25 million frames and approximately 200TB of passband IQ data. This dataset features comprehensive ground-truth metadata aligned with SigMF principles~\cite{SigMFStandardRef}. Furthermore, to facilitate contemporary machine learning workflows, particularly those employing vision-based techniques, we provide methodologies and derived annotations in the COCO format~\cite{Lin2014MicrosoftCOCO} for spectrogram representations of the data. This enables the application of object detection models for sophisticated time-frequency signal analysis, such as localization and classification. The CSRD2025 specification includes standardized 8:1:1 training, validation, and test splits (defined via frame indices) to promote reproducible research and ensure consistent benchmarking across studies.

The resources presented herein aim to significantly lower the barrier for training and evaluating LAMs in wireless communications, thereby accelerating research progress in AI-driven spectrum sensing and intelligent spectrum management. The remainder of this paper is structured as follows: Section~\ref{sec:challenges_related_work} further discusses the challenges of spectrum data acquisition and reviews related work. Section~\ref{sec:csrd_framework} provides a detailed exposition of the CSRD simulation framework. Section~\ref{sec:dataset} comprehensively characterizes the CSRD2025 dataset and its structure. Finally, Section~\ref{sec:conclusion} summarizes the work and outlines future directions.

%======================================================================
% SECTION II: Challenges and Related Work
%======================================================================
\section{Spectrum Data Challenges and Related Work}
\label{sec:challenges_related_work}
The effective application of Large AI Models (LAMs) to complex spectrum sensing tasks is fundamentally contingent upon the availability of extensive, diverse, and accurately annotated training data. However, the acquisition of such data in the RF domain presents a unique confluence of challenges that are considerably more pronounced than in fields like computer vision or natural language processing. This section first elaborates on these intrinsic difficulties, particularly those associated with Over-the-Air (OTA) data collection, and subsequently reviews existing spectrum datasets and generation methodologies in the context of these challenges and the demanding requirements of LAMs.

\subsection{The Intrinsic Difficulties of Spectrum Data Acquisition}
The aspiration to train LAMs directly on the rich tapestry of signals present in real-world RF environments encounters several deeply rooted obstacles:

\textbf{1) High Cost and Complexity of Sensing Infrastructure:} Unlike ubiquitous cameras or readily available text corpora, capturing the RF spectrum across relevant frequency bands requires specialized and expensive equipment. This includes calibrated antenna systems, high-dynamic-range spectrum analyzers, or research-grade Software-Defined Radios (SDRs), representing significant capital investment~\cite{2025keysight, brand_ettus_nodate}. Comprehensive multi-band or multi-node collection campaigns quickly become financially prohibitive for many research groups.

\textbf{2) Requirement for Niche Expertise:} Meaningful RF data acquisition necessitates deep domain knowledge in radio propagation, spectrum regulations, instrumentation settings (sampling rates, filtering, gain control), and calibration procedures. Incorrect setup can easily corrupt data in subtle ways. This expertise barrier prevents large-scale data collection through non-specialist means like crowdsourcing, common in CV/NLP.

\textbf{3) Environmental Dynamism and Regulatory Constraints:} The RF spectrum is inherently non-stationary; signal presence, power levels, and interference patterns fluctuate unpredictably over time and location. Capturing representative diversity demands extensive, long-term monitoring across various environments (urban, rural, indoor, outdoor), further increasing costs. Additionally, monitoring certain frequency bands is subject to strict regulations and licensing, potentially limiting the scope of OTA measurements.

\begin{figure}[t]
  \centering
  \includegraphics[width=0.9\linewidth]{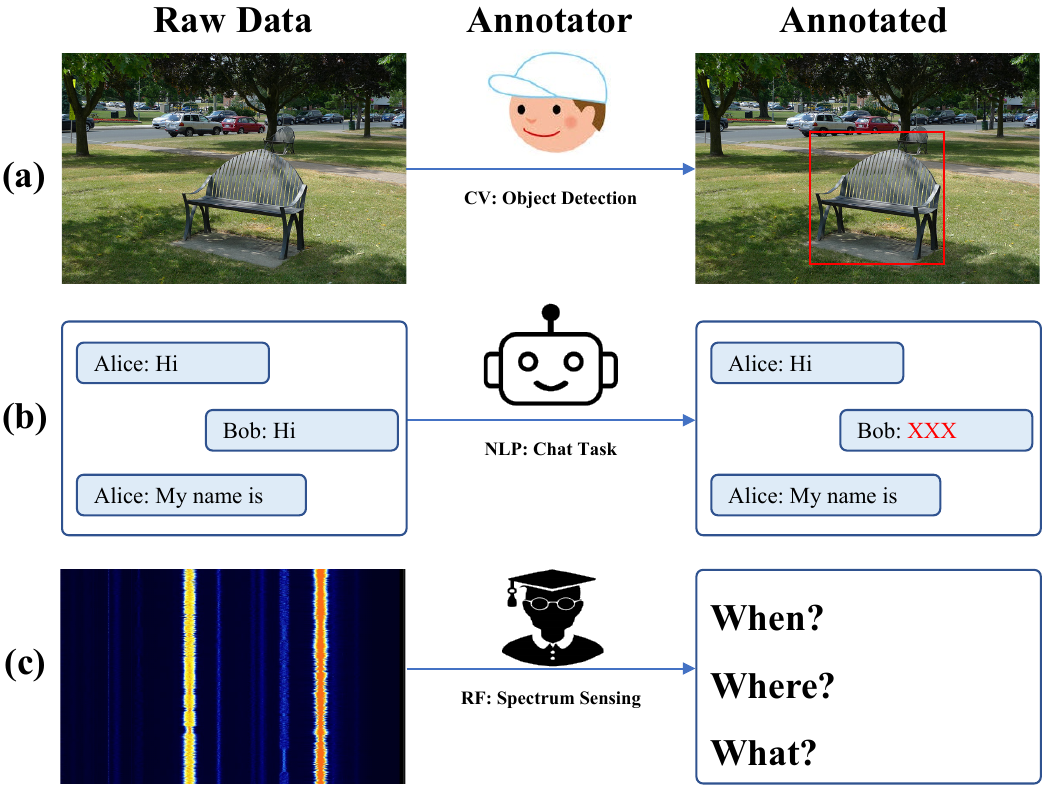}
  \caption{Annotation complexity comparison: (a) Object detection in CV often involves intuitive bounding box drawing. (b) Masked language modeling in NLP can leverage automated annotation. (c) Spectrum sensing requires expert inference from spectrograms or other representations to determine signal parameters (\textit{when}, \textit{where} in frequency, \textit{what} type) which are not directly perceptible.}
  \label{fig:annotation_challenges}
\end{figure}

\textbf{4) The Annotation Impasse (Especially for OTA Data):} Assigning accurate, detailed labels to RF data is arguably a greater bottleneck than acquisition itself, particularly for OTA data needed for supervised LAM training (Fig.\ref{fig:annotation_challenges}).
\begin{itemize}
  \item \textit{Expert Inference vs. Direct Perception:} Key signal parameters (modulation type, bandwidth, center frequency, duration, protocol) are not directly perceptible like visual objects. They must be inferred through complex signal analysis (e.g., spectral, cyclostationary), demanding annotators with significant signal processing expertise.
  \item \textit{Ambiguity and Overlap:} Real-world spectra often contain weak, noisy, transient, or overlapping signals. Disentangling these components and assigning precise time-frequency boundaries and parameters is extremely difficult and often ambiguous, even for experts, hindering the creation of clean labels.
  \item \textit{Lack of Scalable Tools:} RF annotation often relies on specialized, manual, or semi-manual tools. There are no widely available equivalents to the rapid annotation tools or crowdsourcing platforms used effectively in CV/NLP\cite{Nguyen2024WRISTWidebandRealTime, Wood2022DeepLearningObject}.
\end{itemize}
These combined challenges make acquiring massive, diverse, and accurately labeled real-world spectrum datasets, analogous to ImageNet or large text corpora, currently intractable for widespread LAM development in the wireless domain.

\subsection{Existing Spectrum Data Acquisition Approaches and Datasets}
In light of the aforementioned difficulties, particularly those associated with large-scale OTA collection, various alternative or complementary data acquisition strategies and dataset generation efforts have been pursued, each presenting its own set of trade-offs concerning realism, scalability, cost, and annotation fidelity:

\textbf{1) Synthetic Data Generation:} Simulation tools (e.g., MATLAB, GNU Radio) offer scalability, cost-effectiveness, perfect ground-truth labels, and precise control over parameters~\cite{OShea2017LearningRobustGeneral, West2021AWidebandSignal, Li2023BoostSP, Vagollari2021JointDetectionand, Hao2024VSLMVirtualSignal, Huang2022RadDetAWideband, Dong2023Intelligentdetection}. Researchers can generate datasets covering diverse modulations, channel models, SNRs, and interference scenarios. Initiatives like VSLM~\cite{Hao2024VSLMVirtualSignal} explore generating "virtual" signals to improve robustness. However, the primary limitation is the \textit{Sim2Real gap}: models trained purely on synthetic data may not generalize well to the unmodeled nuances of real RF environments~\cite{Li2023BoostSP}. Accurately simulating the full complexity of real-world propagation and hardware effects remains challenging.

\textbf{2) Hardware-in-the-Loop (HIL) Emulation:} Platforms like Colosseum~\cite{Bonati2021ColosseumLargeScale} bridge the Sim2Real gap by using real SDRs interacting within a large-scale channel emulator. This incorporates actual hardware behavior while maintaining controlled, repeatable propagation environments. HIL allows testing full protocol stacks under complex interference. However, such platforms are expensive, limiting accessibility, and the channel models, while sophisticated, are still emulations. Datasets generated depend on specific user experiments.

\textbf{3) Real-World (OTA) Data Acquisition:} Collecting data OTA provides maximum realism, capturing genuine channel, interference, and hardware effects~\cite{OShea2018OvertheAirDeep, Morehouse2023AnOptimizedFaster}. Public datasets like RadioML 2018.01A~\cite{OShea2018OvertheAirDeep}, SPREAD~\cite{Wicht2022SpectrogramDataSet, Nguyen2024WRISTWidebandRealTime}, Drone RF~\cite{Basak2022CombinedRFBased}, or data from platforms like Electrosense~\cite{Scalingi2023AFrameworkfor} offer valuable real-world snapshots. However, as discussed in Section~\ref{sec:challenges_related_work}.A, OTA collection is costly, hard to scale, difficult to reproduce, and severely limited by annotation challenges, especially for the fine-grained labels needed for tasks beyond simple classification~\cite{Nguyen2024WRISTWidebandRealTime}.

\subsection{The Dataset Deficit for Spectrum LAMs}
\label{subsec:dataset_deficit}
The aforementioned challenges in spectrum data acquisition become particularly acute when considering the demanding requirements of LAMs. An evaluation of the current landscape of publicly available RF datasets against these needs—massive volume, extensive diversity across numerous parameters, and detailed, accurate labels—reveals a significant \textit{dataset deficit}. Specifically, existing public datasets are often orders of magnitude smaller in \textbf{scale} than those successfully employed in mainstream AI fields like CV and NLP. Furthermore, their \textbf{diversity} concerning complex interference scenarios, varied RF hardware impairments, an extensive range of modern and cutting-edge modulation schemes, and diverse channel conditions frequently remains limited, with many datasets tailored to specific, relatively narrow applications (e.g., AMC using a few modulations, or radar detection~\cite{Soltani2022FindingWaldoin, Huang2022RadDetAWideband}). The inherent difficulty in precisely annotating real-world OTA data severely restricts the availability of large-scale datasets with the required \textbf{annotation quality and granularity} suitable for supervised learning tasks that demand fine-grained information, such as precise time-frequency localization or detailed parameter estimation of multiple concurrent signals. Compounding these issues, inconsistencies in data formats, metadata structures, and evaluation protocols across various datasets hinder effective benchmarking, meta-learning, and collaborative advancement due to a lack of \textbf{standardization}.

This pronounced dataset deficit forms a critical bottleneck, significantly impeding the development, rigorous training, and comprehensive evaluation of robust and generalizable LAMs for advanced spectrum sensing tasks. Our work directly addresses this challenge by proposing the ChangShuoRadioData (CSRD) framework. As detailed in Section~\ref{sec:csrd_framework}, CSRD is a scalable, open-source platform for generating large, diverse, and high-fidelity \textit{synthetic} datasets, exemplified by the CSRD2025 instance (Section~\ref{sec:dataset}). By providing perfect ground-truth metadata, including annotations suitable for complex tasks like time-frequency object detection, our approach aims to fulfill the critical data requirements of the synthetic data generation stage (Stage 1) outlined in our proposed strategy (Fig.~\ref{fig:three_stage}), thereby lowering a key barrier to LAM research in the wireless domain.

\begin{figure*}[t]
  \centering
  \includegraphics[width=0.8\linewidth]{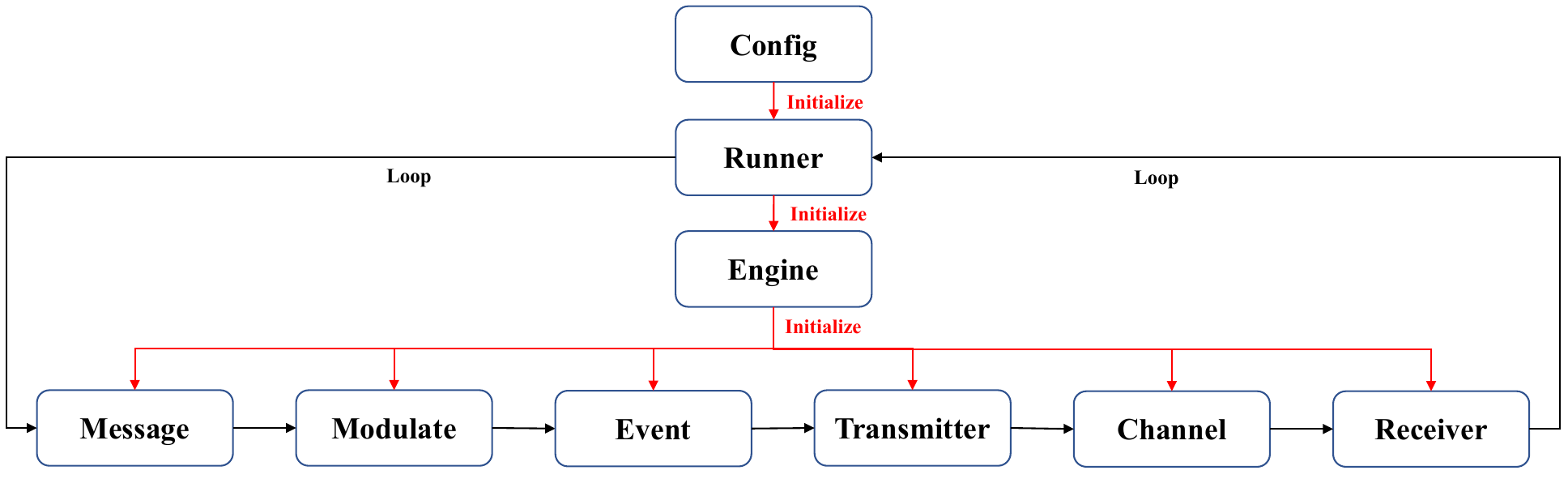}
  \caption{Overview of the CSRD framework architecture, illustrating the modular components and simulation workflow for generating large-scale synthetic spectrum datasets.}
  \label{fig:csrd_framework}
\end{figure*}
%======================================================================
% SECTION IV: ChangShuoRadioData FRAMEWORK
%======================================================================
\section{ChangShuoRadioData: A Scalable Framework for High-Fidelity Synthetic Spectrum Data Generation}
\label{sec:csrd_framework}
Addressing the limitations of existing data sources and the impracticality of large-scale real-world collection (Section~\ref{sec:challenges_related_work}), we developed the ChangShuoRadioData (CSRD) framework. CSRD is an open-source simulation platform engineered for the systematic generation of large-scale, high-fidelity synthetic datasets tailored for spectrum-aware LAM development. It serves as the cornerstone of the synthetic data stage (Stage 1) in our proposed strategy (Fig.~\ref{fig:three_stage}), providing a controlled yet realistic virtual environment for generating data with perfect ground truth. This section details the CSRD framework's architecture, core modeling components, and capabilities. Figure~\ref{fig:csrd_framework} provides a conceptual overview.

\subsection{Architectural Principles and Simulation Workflow}
\label{subsec:framework_architecture_revised}
The CSRD framework is built upon key principles of \textbf{Modularity}, allowing for the independent development and substitution of components such as modulation schemes or channel models; \textbf{Flexibility}, achieved through a configuration-driven approach where users define simulation scenarios via external JSON parameter files; \textbf{Realism}, by incorporating detailed mathematical models of physical phenomena, including channel effects and critical RF hardware impairments; and \textbf{Scalability}, supporting efficient computation and parallel processing for the generation of massive datasets.

The framework simulates the complete end-to-end signal chain, from source information generation, through modulation and transmitter-induced distortions, signal propagation via diverse wireless channel models, signal aggregation at the receiver, to the introduction of receiver-side noise and impairments. The core simulation engine, orchestrated by a dedicated runner module, operates on a frame-by-frame basis. For each simulated frame, macroscopic scenario parameters (e.g., number of active transmitters and receivers, types of signals, choice of channel model) are stochastically sampled from distributions defined in the external configuration files. Subsequently, the engine dynamically instantiates the required simulation modules (e.g., modulator, channel model, impairment models) with parameters specific to that frame. Signals are then synthesized, scheduled in time and frequency by an event controller (which can introduce controlled spectral overlap to simulate interference), propagated through the selected channel model, aggregated at one or more simulated receivers, combined with receiver-specific noise and impairments, and finally archived along with comprehensive metadata by the runner module. This cycle, designed for reproducibility and parallel execution, is repeated to construct the large-scale dataset. The decoupling of simulation logic from experimental setup via the configuration-driven architecture is paramount for fostering reproducibility and simplifying the exploration of diverse operational scenarios.

\subsection{Core Components and Modeling Fidelity}
\label{subsec:core_components_revised}
The framework's ability to generate diverse, high-fidelity data relies on the detailed implementation of its modular components.

\textbf{1) Configuration System:} Utilizes hierarchical JSON files for specifying global and component-specific parameters, enabling flexible scenario definition and reproducibility without code modification.

\textbf{2) Simulation Runner:} Manages large-scale simulation execution, including initialization, parallel processing orchestration (leveraging MATLAB's parallel features), standardized data archiving (`.mat' for IQ, `.json' for metadata), and progress monitoring.

\textbf{3) Core Simulation Engine:} Dynamically translates configurations into simulation instances per frame by stochastically sampling parameters, configuring modules contextually, instantiating necessary objects, and orchestrating the sequential execution of the simulation pipeline.

\textbf{4) High-Fidelity Simulation Pipeline modules:} The scientific fidelity of the generated data stems from the detailed physical and hardware models within the pipeline modules:

\begin{itemize}
  \item \textbf{Source Message Modeling:} This initial stage generates the baseband information sequence. For digital systems, pseudo-random binary sequences are typically produced. To model analog transmissions (e.g., AM/FM) or specific interference types, the framework can also load source content, such as audio waveforms, from external files, allowing for the simulation of heterogeneous signal environments.

  \item \textbf{Comprehensive Modulation Scheme Emulation:}  A cornerstone of the framework, this module provides an extensive and configurable library of modulation techniques, critical for representing diverse spectrum usage. It encompasses analog formats (AM variants including DSB, SSB, VSB; FM; PM) and a wide array of digital single-carrier schemes. These include fundamental types like Amplitude Shift Keying (ASK/OOK), various orders of Phase Shift Keying (PSK, e.g., BPSK, QPSK, up to 64-PSK) and Quadrature Amplitude Modulation (QAM, e.g., 16-QAM up to 4096-QAM, including MIL-STD-188 variants), Frequency Shift Keying (FSK), and diverse Continuous Phase Modulation (CPM) types such as GMSK, MSK, and CPFSK. Importantly, the framework also implements modern multi-carrier techniques like Orthogonal Frequency Division Multiplexing (OFDM) and Single-Carrier Frequency Division Multiple Access (SCFDMA), along with the advanced delay-Doppler domain waveform, Orthogonal Time Frequency Space (OTFS). Essential practical details, such as randomized Root-Raised Cosine (RRC) pulse shaping for spectral containment and Orthogonal Space-Time Block Coding (OSTBC) for MIMO transmissions, are incorporated, contributing significantly to the realism and diversity of the generated signals (approximately 100 distinct types configurable, see Section~\ref{subsec:dataset_cocometa}).

  \item \textbf{Event Management and Spectral Coexistence Simulation:} This module orchestrates the spatio-temporal placement of multiple signal segments within each simulation frame to create realistic spectral scenarios. It manages temporal dynamics by assigning start times and randomized idle periods between segments from the same source, enabling the simulation of both continuous and bursty traffic patterns characteristic of many communication protocols. Concurrently, it performs spectral allocation, distributing active transmitters across the simulated frequency band with consideration for their bandwidths and optional guard bands. A key capability is the controlled simulation of interference: users can configure the probability and extent of spectral overlap between concurrently active signals, allowing the generation of scenarios ranging from minimal interaction to significant co-channel or adjacent-channel interference.

  \item \textbf{Transmitter RF Front-End Impairment Modeling:} To enhance simulation realism and bridge the Sim2Real gap, this module introduces critical non-ideal hardware effects commonly found in transmitter chains. These include IQ Imbalance, resulting from gain and phase mismatches in the I/Q paths; DC Offset, which can introduce spurious tones; Phase Noise, stemming from local oscillator instabilities and modeled using spectral masks; and Power Amplifier (PA) Nonlinearity, a significant source of distortion. The framework employs flexible behavioral models for PAs (e.g., Cubic polynomial, Saleh, Rapp, accessible via \texttt{comm.MemorylessNonlinearity}) to capture effects such as gain compression and spectral regrowth (AM/AM and AM/PM conversion). The type and severity of these impairments are stochastically varied based on configuration files, representing a diverse range of transmitter hardware quality.

  \item \textbf{Wireless Channel Emulation:} This module simulates the signal's transformation during propagation. It combines standard statistical models with advanced site-specific simulation.
        \begin{itemize}
          \item \textit{Path Loss and Statistical Fading:} Fundamental distance-based path loss is applied, optionally augmented by atmospheric attenuation. Standard statistical multipath fading models, such as Rayleigh and Rician (with configurable K-factor, path delays, gains, etc.), are implemented using MATLAB's Communications Toolbox functionalities. Doppler shift based on simulated mobility introduces channel time variation.

          \item \textit{Site-Specific Ray Tracing:} For enhanced environmental fidelity, the framework integrates an optional ray tracing propagation model. This model leverages 3D environmental geometry, primarily building structures, sourced from OpenStreetMap (OSM) data.
                OSM map files corresponding to 25 distinct geographic locations are utilized (e.g., dense urban high-rise, urban canyon, sparse suburban, industrial park, university campus, rural village center, open farmland, desert, coastal town), downloading 10 representative 2km x 2km examples for each category. Figure~\ref{fig:osm_examples} illustrates the diversity of these environments, ensuring simulations can represent propagation in vastly different settings.

                \begin{figure*}[ht]
                  \centering
                  \begin{subfigure}[b]{0.245\linewidth}
                    \includegraphics[width=\textwidth]{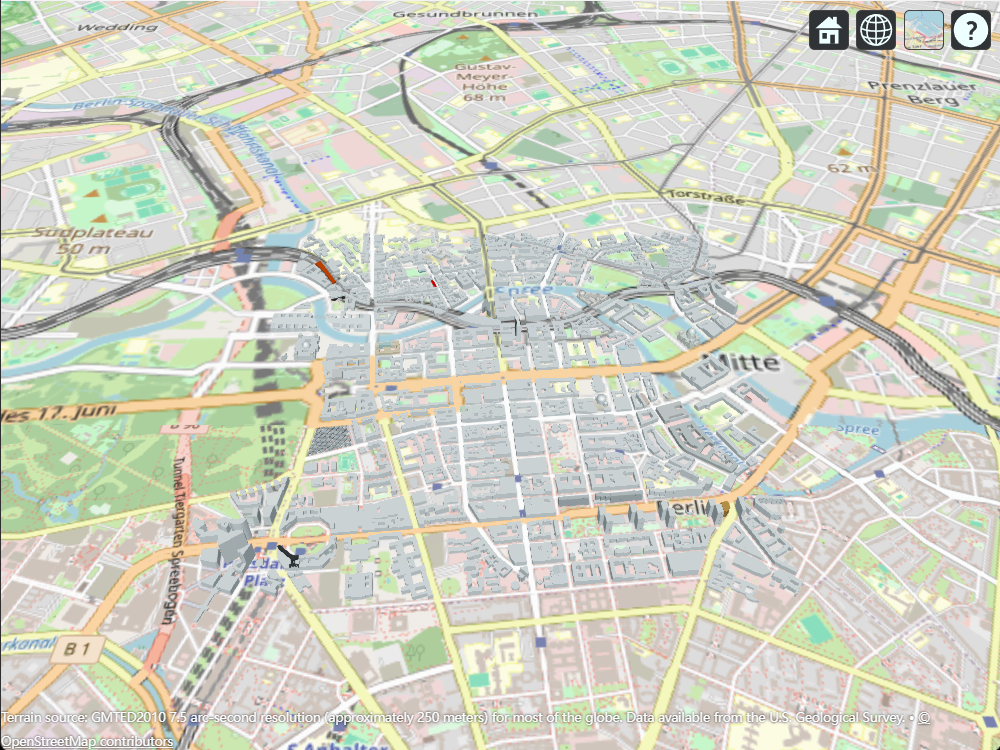}
                    \caption{Dense Urban}
                  \end{subfigure}\hfill
                  \begin{subfigure}[b]{0.245\linewidth}
                    \includegraphics[width=\textwidth]{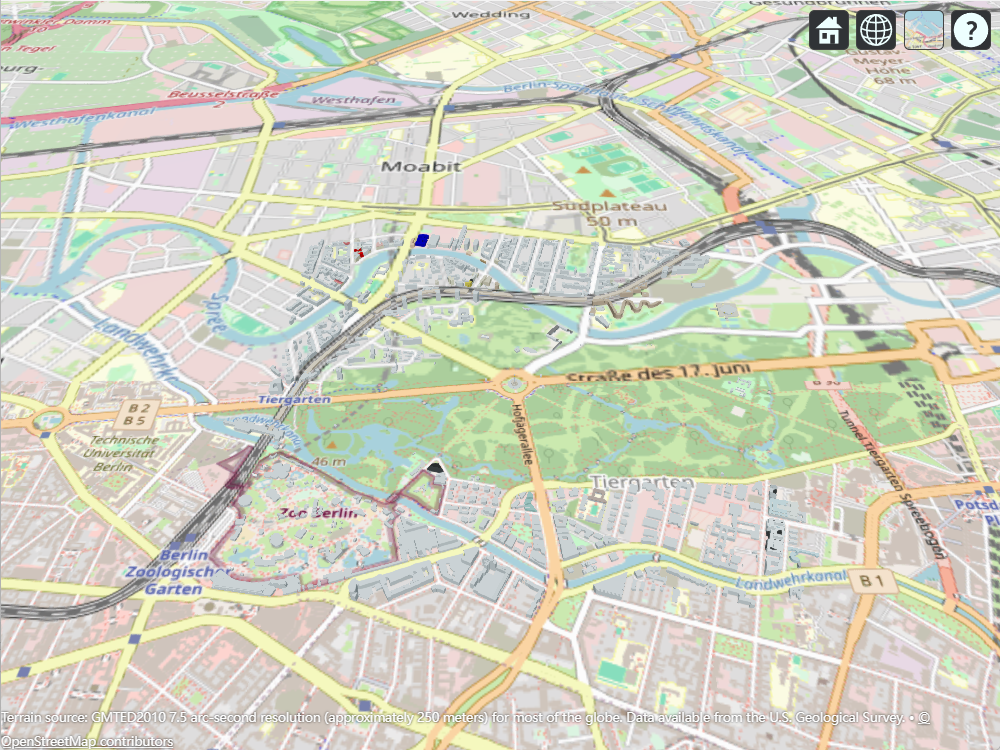}
                    \caption{Park}
                  \end{subfigure}\hfill
                  \begin{subfigure}[b]{0.245\linewidth}
                    \includegraphics[width=\textwidth]{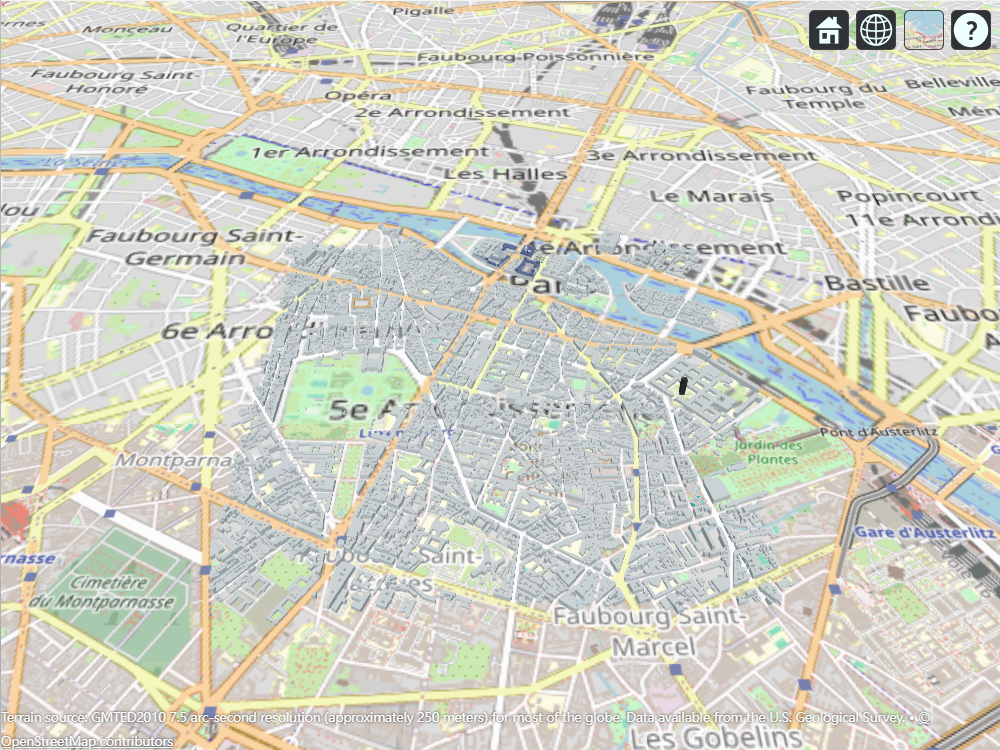}
                    \caption{Campus}
                  \end{subfigure}\hfill
                  \begin{subfigure}[b]{0.245\linewidth}
                    \includegraphics[width=\textwidth]{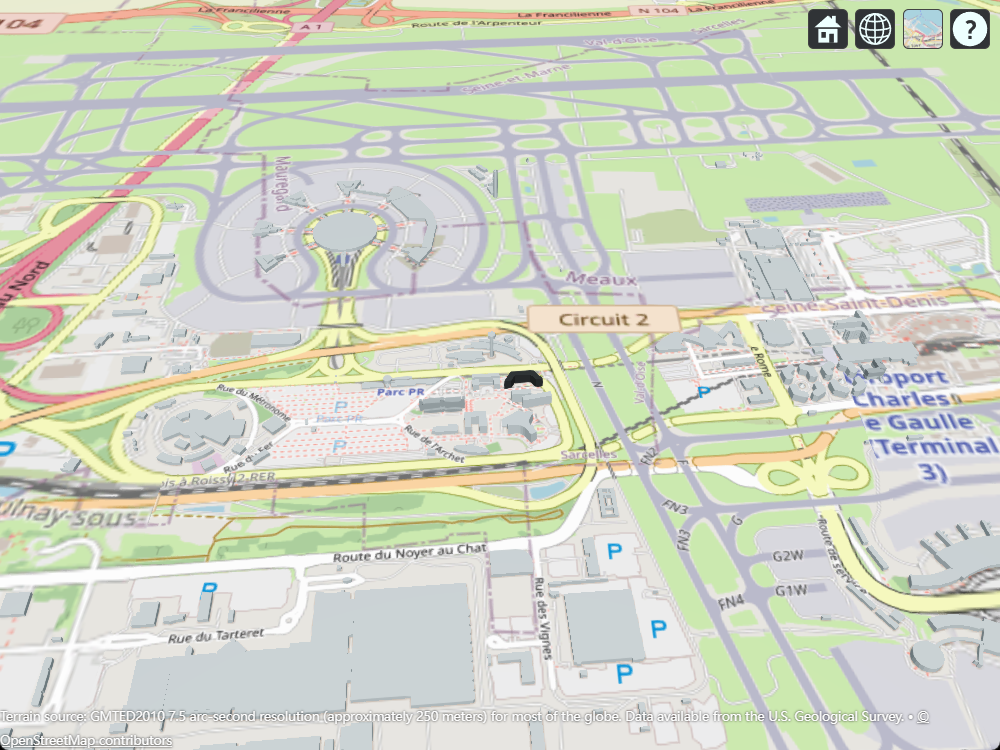}
                    \caption{Airport}
                  \end{subfigure}

                  \vspace{0.5em}

                  \begin{subfigure}[b]{0.245\linewidth}
                    \includegraphics[width=\textwidth]{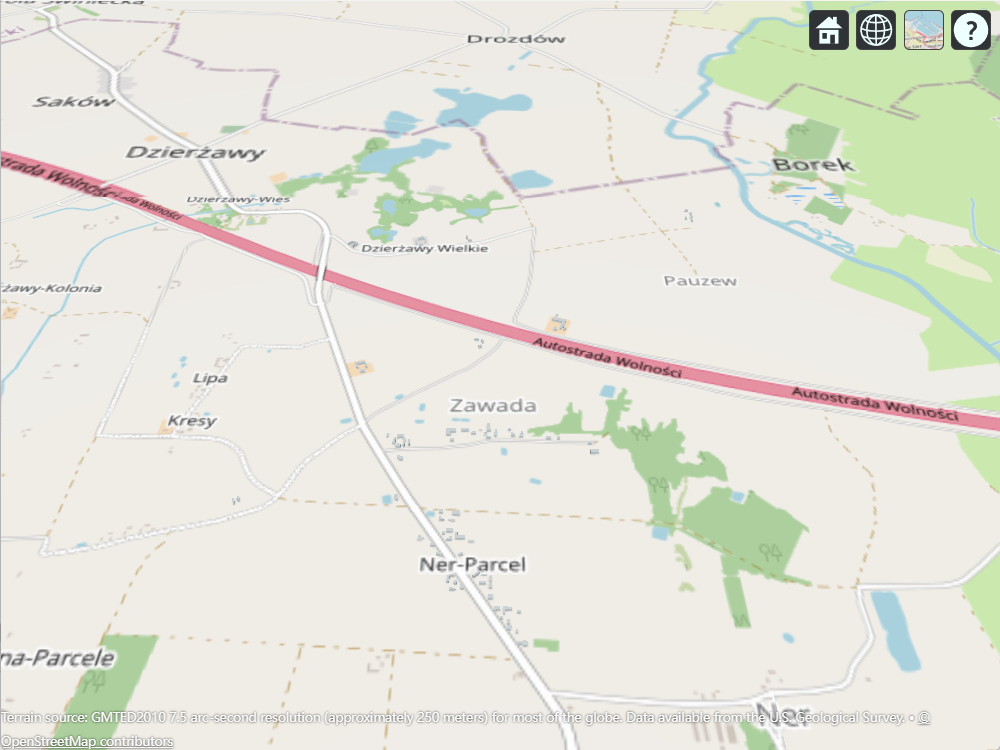}
                    \caption{Farmland}
                  \end{subfigure}\hfill
                  \begin{subfigure}[b]{0.245\linewidth}
                    \includegraphics[width=\textwidth]{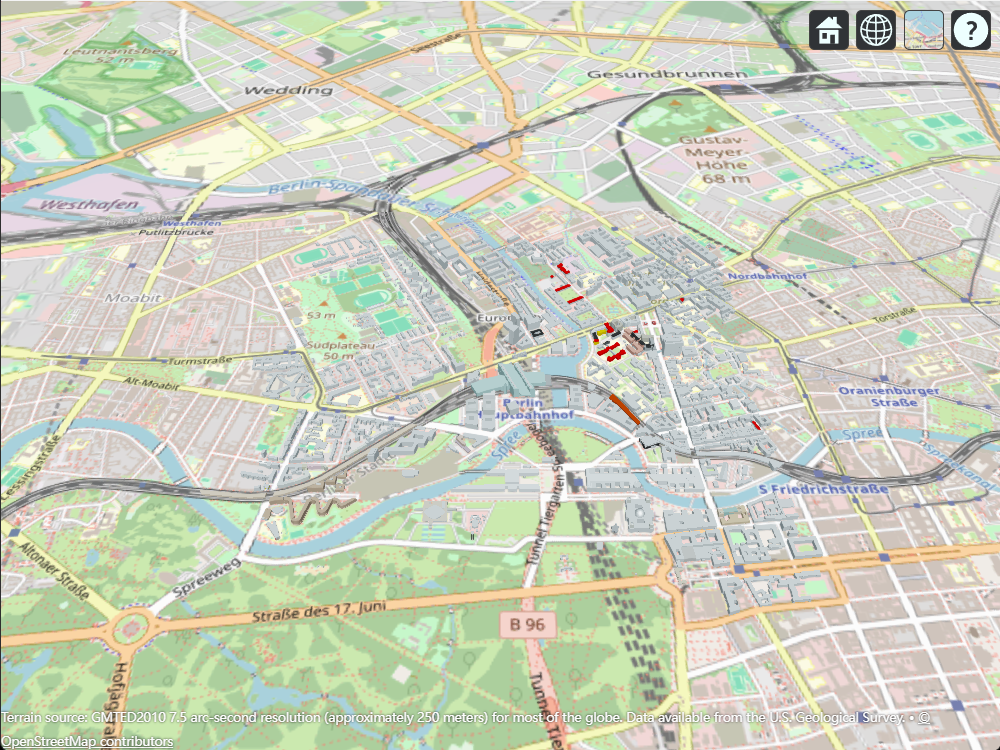}
                    \caption{Train Station}
                  \end{subfigure}\hfill
                  \begin{subfigure}[b]{0.245\linewidth}
                    \includegraphics[width=\textwidth]{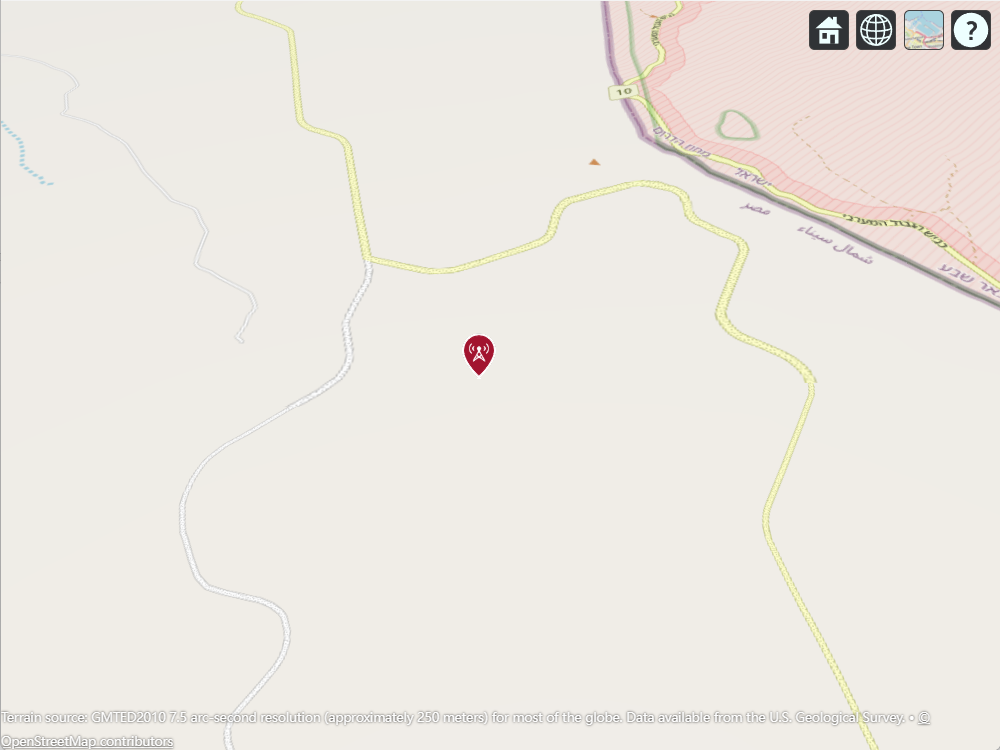}
                    \caption{Desert}
                  \end{subfigure}\hfill
                  \begin{subfigure}[b]{0.245\linewidth}
                    \includegraphics[width=\textwidth]{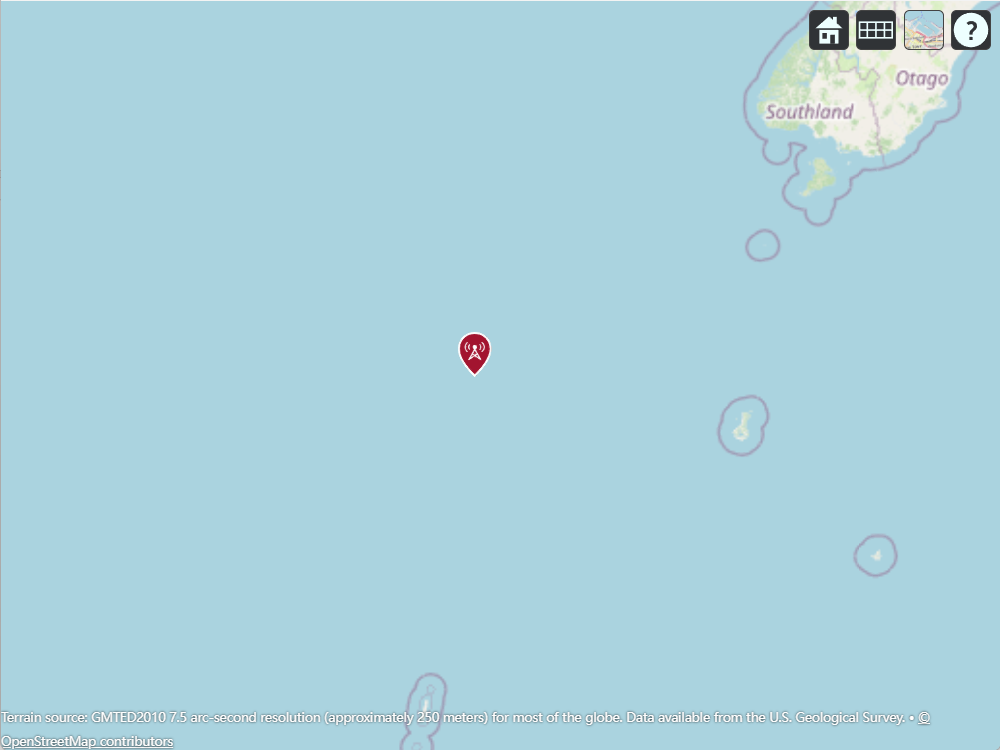}
                    \caption{Ocean}
                  \end{subfigure}
                  \caption{Examples of diverse geographical environments sourced from OpenStreetMap data. The eight scenarios depicted are randomly selected from 25 distinct OpenStreetMap environment types and used for ray tracing simulations.}
                  \label{fig:osm_examples}
                \end{figure*}

                The ray tracing engine within CSRD processes the building and terrain data from the loaded  `.osm' file. It then computes propagation paths between specified transmitter and receiver locations using techniques like Shooting and Bouncing Rays (SBR) or the Image Method, considering reflections and diffractions off structures, governed by configurable parameters such as maximum interaction counts. Figure~\ref{fig:raytracing_simulation_examples} visualizes typical ray tracing results in Barcelonès.
                The outcome is a detailed, site-specific channel impulse response incorporating path losses, delays, phase shifts, and angle-of-arrival/departure information for dominant rays. This complex channel response is then applied to the simulated signal, providing a higher level of realism compared to purely statistical models, especially in complex urban or indoor environments.

          \item \textit{Antenna Architecture Support:} The channel modeling correctly simulates SISO, MISO, SIMO, and MIMO configurations based on the number of antennas defined for the transmitter and receiver.
        \end{itemize}

  \item \textbf{Receiver RF Front-End Impairment Modeling:} This final stage introduces receiver-induced noise and imperfections. Additive White Gaussian Noise (AWGN) is added based on a randomized receiver Noise Figure, establishing the fundamental noise floor. Analogous to the transmitter, effects like IQ imbalance, DC offset, and Low-Noise Amplifier (LNA) nonlinearity are simulated with stochastically chosen parameters. The module calculates and records the ground-truth SNR for each signal relative to the thermal noise before other receiver impairments are applied, providing a crucial label for supervised learning.
\end{itemize}

\begin{figure}[t]
  \centering
  \begin{subfigure}[b]{0.49\linewidth}
    \includegraphics[width=\textwidth]{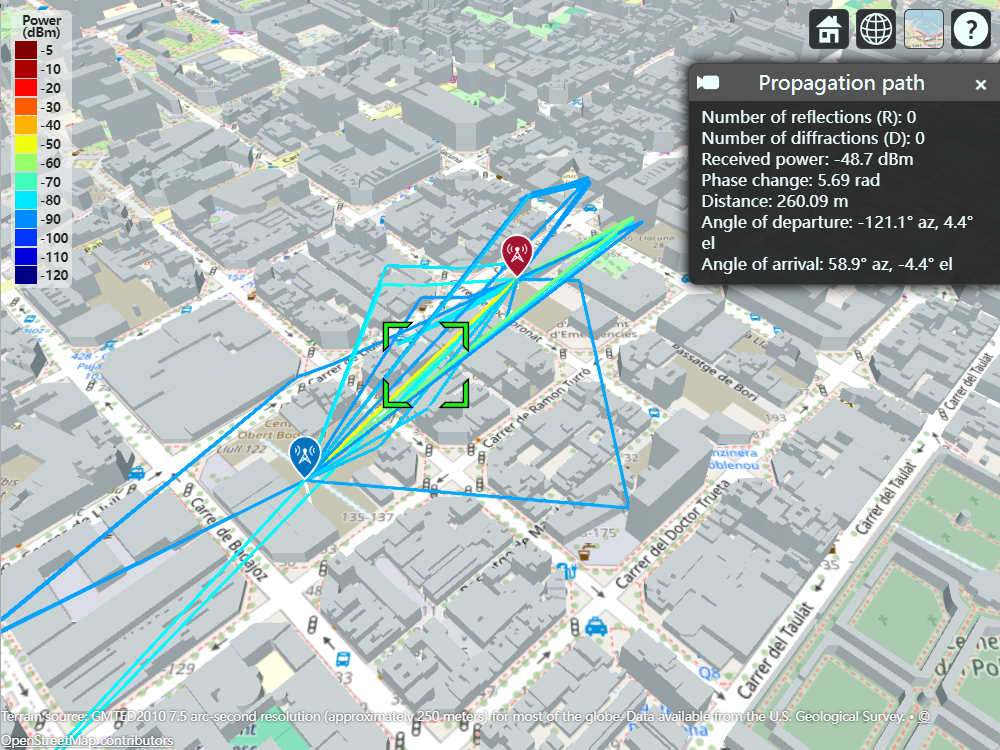}
    \caption{Ray tracing paths between Tx and Rx}
    \label{fig:ray_paths}
  \end{subfigure}\hfill
  \begin{subfigure}[b]{0.49\linewidth}
    \includegraphics[width=\textwidth]{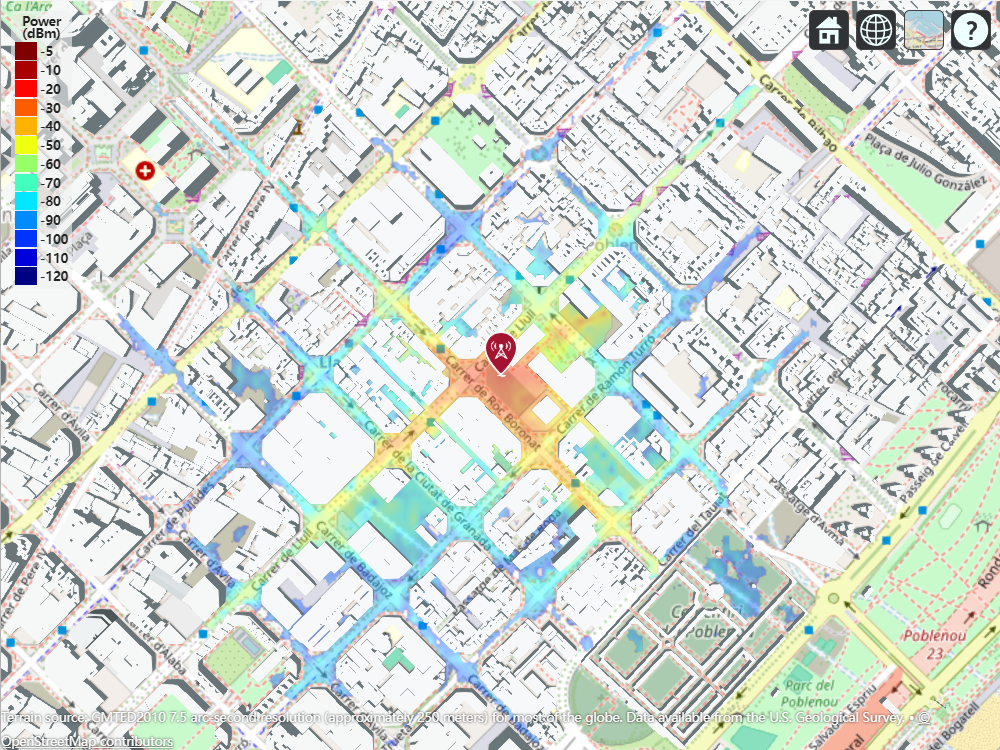}
    \caption{Signal coverage map for the transmitter}
    \label{fig:coverage_map}
  \end{subfigure}
  \caption{Examples of ray tracing simulation outputs in Barcelonès. (a) Computed propagation paths (rays) between a transmitter (Tx) and a receiver (Rx), considering reflections and diffractions using building geometry from OpenStreetMap (OSM). (b) Corresponding signal coverage map for the transmitter in the same environment.}
  \label{fig:raytracing_simulation_examples}
\end{figure}

\subsection{Framework Scalability and Extensibility}
\label{subsec:scalability_extensibility_revised}
CSRD is designed for large-scale generation and community adaptation. Its computational scalability is enabled by efficient implementation and parallel processing support. The modular architecture and configuration system ensure extensibility, allowing easy addition of new components or modification of parameters without altering core code. The framework's generation capacity is limited primarily by computational resources, allowing users to create datasets far larger than the specific CSRD2025 instance presented here.

In summary, the CSRD framework provides a powerful, flexible, and scalable open-source tool for generating the extensive, diverse, and realistic synthetic datasets crucial for advancing research in AI-driven spectrum sensing and management.

\begin{table*}[t]
  \centering
  \caption{Comparison of CSRD2025 with Selected Public RF Datasets.}
  \label{tab:dataset_comparison}
  \begin{tabularx}{\linewidth}{llllll}
    \toprule
    Dataset                                                & Size                             & Mod Classes         & Frame Number & Frame Length                      & Notes                                                               \\
    \midrule
    \textbf{CSRD2025 (Ours)}                               & \textbf{\textasciitilde{}200 TB} & \textbf{100}        & 25M          & $[2K, 4M]$                        & Passband, \textbf{High diversity}                                   \\
    SPREAD~\cite{Wicht2022SpectrogramDataSet}              & \textasciitilde{}8.1 TB          & \textasciitilde{}10 & 0.02M        & \{112500, 202500, 27000, 562500\} & Passband, Spectrum monitoring focus                                 \\
    TorchSig Sig53~\cite{DatasetSig53Ref}                  & \textasciitilde{}1.4 TB          & 53                  & 0.55M        & 262144                            & Passband, Wideband Operations focus                                 \\
    RadioML 2018.01A~\cite{OShea2018OvertheAirDeep}        & \textasciitilde{}18 GB           & 24                  & 2.55M        & 1024                              & Baseband, Widely used for AMC                                       \\
    HisarMod~\cite{DatasetHisarModRef}                     & \textasciitilde{}5.12 GB         & 26                  & 0.78M        & 1024                              & Baseband, Focused on AMC                                            \\
    RML22~\cite{Sathyanarayanan2023RML22}                  & $<$ 1 GB                         & 10                  & 1.4M         & 128                               & Baseband, Improved Version of~\cite{OShea2017LearningRobustGeneral} \\
    RadioML 2016.10A~\cite{OShea2017LearningRobustGeneral} & $<$ 1 GB                         & 11                  & 0.22M        & 128                               & Baseband, Early benchmark for AMC                                   \\
    \bottomrule
  \end{tabularx}
\end{table*}

%======================================================================
% SECTION V: The CSRD2025 Dataset
%======================================================================
\section{The CSRD2025 Dataset: Characteristics, Structure, and Access}
\label{sec:dataset}
Using the CSRD framework (Section~\ref{sec:csrd_framework}) with its reference configuration and significant computational resources, we generated the CSRD2025 dataset. This dataset instance is a large-scale, high-fidelity synthetic resource specifically designed to facilitate the development and evaluation of spectrum-aware AI models, especially LAMs. This section details the characteristics of CSRD2025, including its scale, diversity, data formats, annotation structure (including COCO for object detection), statistical properties, and access guidelines.

\subsection{Dataset Scale and Comparison}
\label{subsec:dataset_scale_compare}
A primary objective in creating CSRD2025 was to provide data at a scale commensurate with the demands of contemporary deep learning models. The CSRD2025 instance characterized in this work was generated from approximately \textbf{10,000,000 distinct simulation scenarios}. These scenarios were designed with diverse channel modeling approaches: approximately 90\% utilized statistical channel models (Rayleigh, Rician), while the remaining 10\% employed the more computationally intensive site-specific ray tracing models based on OSM data (as detailed in Section~\ref{sec:csrd_framework}).

Each individual scenario simulation involved a varying number of active transmitters (1 to 4) and receivers (1 to 4). Each active receiver within a scenario captures a single data frame. The total number of recorded frames ($N_{\text{frames}}$) in the dataset is thus the sum of frames generated across all scenarios ($N_{\text{scenarios}}$), considering the number of active receivers ($N_{Rx,i}$) in each scenario $i$:
\begin{equation}
  N_{\text{frames}} = \sum_{i=1}^{N_{\text{scenarios}}} N_{Rx,i}
  \label{eq:total_frames}
\end{equation}

This generation process resulted in the following quantitative characteristics for CSRD2025:
\begin{itemize}
  \item \textbf{Total Recorded Frames:} Over \textbf{25,000,000 (25M) unique recorded frames}, derived from the aforementioned scenarios, with each frame representing a unique receiver's perspective within its scenario.
  \item \textbf{Data Size:} Approximately \textbf{200 Terabytes (TB)} of complex passband IQ data and associated metadata. This represents a significant increase in scale compared to seminal datasets for RF machine learning; for instance, it is approximately \textbf{10,000 times larger} than the widely used 18GB RadioML 2018.01A dataset~\cite{OShea2018OvertheAirDeep}.
  \item \textbf{Frame Length:} CSRD2025 features variable frame lengths, typically ranging from \textbf{2,000 to 4,000,000 samples per frame}, reflecting the diverse durations of real-world signal transmissions and the number of signal segments within each frame.
\end{itemize}
This massive scale, particularly in terms of total data volume and the number of individual recorded frames, alongside the diversity discussed subsequently, significantly surpasses many existing public datasets. Table~\ref{tab:dataset_comparison} provides a comparative overview of CSRD2025 against several other notable RF datasets.

As highlighted and further detailed in Table~\ref{tab:dataset_comparison}, CSRD2025 distinguishes itself not only by its sheer volume—being orders of magnitude larger than foundational datasets like RadioML 2018.01A—but also by its high number of modulation classes (100 distinct types) and the generation of passband signals, which are more representative of signals captured by an antenna before down-conversion. The variable frame length is another key feature, designed to better emulate the varying durations of signal activity one might encounter in realistic spectrum monitoring. In contrast, many existing datasets, particularly those focused on Automatic Modulation Classification (AMC), often utilize shorter, fixed-length baseband signals. While datasets like SPREAD and TorchSig Sig53 offer larger volumes of passband data compared to older benchmarks, CSRD2025 aims to provide a significantly larger number of recorded frames derived from diverse scenarios and a broader range of modulation classes, coupled with extensive configurable diversity in signal parameters, channel conditions, and RF impairments (detailed in Section~\ref{subsec:dataset_diversity}). This scale and engineered diversity are crucial for training data-hungry LAMs and exploring their generalization capabilities in complex spectral environments. It is important to reiterate that users can leverage the open-source CSRD framework to generate even larger datasets or datasets with different parameter distributions tailored to specific research needs by adjusting the provided configuration files and allocating sufficient computational resources.

\begin{table*}[t]
  \centering
  \caption{Summary of Key Parameter Variations in the CSRD2025 Dataset (Based on Default Configuration).}
  \label{tab:dataset_params_summary_revised}
  \begin{tabularx}{\linewidth}{llX}
    \toprule
    Parameter Category   & Specific Parameter                 & Range / Choices / Distribution Notes                                                                       \\
    \midrule
    Scenario Complexity  & Number of Transmitters per Frame   & Uniform Discrete [1, 4]                                                                                    \\
                         & Number of Receivers per Frame      & Uniform Discrete [1, 4]                                                                                    \\
                         & Segments per Transmitter           & Uniform Discrete [1, 3]                                                                                    \\ \addlinespace
    Signal Types         & Modulation Families                & Analog (AM, FM, PM), Digital Single-Carrier (ASK, PSK, QAM, FSK, CPM), Multi-Carrier (OFDM, SCFDMA), OTFS  \\
                         & Modulation Classes (Total)         & 100 distinct types (see Section~\ref{subsec:dataset_cocometa})                                             \\
                         & Modulation Orders (Examples)       & PSK (2-64), QAM (8-4096), FSK (2-8), etc.                                                                  \\ \addlinespace
    Signal Parameters    & Symbol Rate                        & Uniform [30 kHz, 50 kHz]                                                                                   \\
                         & Bandwidth                          & Varies (e.g., ~30 kHz to ~150 kHz depending on Mod/SR/Pulse Shape)
    \\
                         & Carrier Frequency                  & Dynamically placed by event manager within simulated band                                                  \\
                         & Signal Duration per Segment        & Varies (e.g., ~500 to ~2000 symbols)
    \\ \addlinespace
    System Config.       & Antenna Architectures (Tx/Rx)      & Uniform Discrete [1, 4] antennas each; supports SISO, MISO, SIMO, MIMO (up to 4x4)                         \\ \addlinespace
    Channel Conditions   & Fading Distributions               & Rayleigh, Rician (K-factor Uniform [1, 9])                                                                 \\
                         & Multipath Profiles                 & 1-3 additional paths, delays logarithmically sampled, gains exponentially decaying                         \\
                         & Max Doppler Shift                  & Corresponds to relative speed Uniform [1.5, 28] m/s                                                        \\
                         & Distance                           & Varied based on Indoor/Outdoor models (e.g., up to 1 km)                                                   \\
                         & Ray Tracing Environments           & Optional selection from 25 OSM-based scene types                                                           \\ \addlinespace
    RF Impairments (Tx)  & PA Nonlinearity Models             & Cubic, Saleh, Ghorbani, Rapp (Uniform selection)                                                           \\
                         & PA Parameter Ranges                & Varied (e.g., IIP3 Uniform [20, 40] dBm)                                                                   \\
                         & Phase Noise Level (@ 10kHz offset) & Uniform [-150, -100] dBc/Hz                                                                                \\
                         & IQ Imbalance (Amp/Phase)           & Uniform [0, 5] dB / Uniform [0, 5] deg                                                                     \\
                         & DC Offset (relative to signal)     & Uniform [-60, -40] dB                                                                                      \\ \addlinespace
    RF Impairments (Rx)  & LNA Nonlinearity Models            & Cubic, Saleh, Ghorbani, Rapp (Uniform selection)                                                           \\
                         & LNA Parameter Ranges               & Varied (similar ranges to Tx PA)                                                                           \\
                         & Noise Figure                       & Uniform [10, 20] dB                                                                                        \\
                         & IQ Imbalance (Amp/Phase)           & Uniform [0, 5] dB / Uniform [0, 5] deg                                                                     \\
                         & DC Offset (relative to signal)     & Uniform [-60, -40] dB                                                                                      \\ \addlinespace
    Received SNR         & Ground Truth SNR per Signal        & Wide distribution resulting from path loss, Tx power, NF (see Fig.~\ref{fig:snr_distribution_placeholder}) \\ \addlinespace
    Spectral Coexistence & Signal Overlap Probability         & 0.15                                                                                                       \\
                         & Signal Overlap Extent              & Uniform [0, 0.15] (0 to 15\% bandwidth overlap)                                                            \\
    \bottomrule
  \end{tabularx}
\end{table*}

\subsection{Engineered Diversity}
\label{subsec:dataset_diversity}
CSRD2025's value stems from its extensive, engineered diversity across multiple dimensions, mirroring real-world spectral heterogeneity. This is achieved through stochastic parameter variation within the CSRD framework based on the reference configuration. Key aspects include:
\begin{itemize}
  \item \textbf{Scenario Complexity:} Varied number of transmitters (1-4), receivers (1-4), and signal segments per transmitter (1-3) per frame.
  \item \textbf{Signal Types:} Comprehensive coverage of 100 distinct modulation types (detailed in Section~\ref{subsec:dataset_cocometa}), spanning analog, digital single-carrier (ASK, PSK, QAM, FSK, CPM), multi-carrier (OFDM, SCFDMA), and OTFS families.
  \item \textbf{Signal Parameters:} Randomized symbol rates (e.g., 30-50 kHz), varying bandwidths (based on modulation, rate, pulse shaping), dynamic carrier frequency placement, and variable segment durations.
  \item \textbf{System Configuration:} Inclusion of SISO, MISO, SIMO, and MIMO (up to 4x4) links.
  \item \textbf{Channel Conditions:} Diverse propagation distances (indoor/outdoor ranges), statistical fading (Rayleigh, Rician with varied K-factors), varied multipath profiles, Doppler shifts (corresponding to speeds 1.5-28 m/s), and optional site-specific ray tracing based on 25 OSM environment types.
  \item \textbf{RF Impairments:} Realistic transmitter (PA nonlinearity, phase noise, IQ imbalance, DC offset) and receiver (LNA nonlinearity, noise figure variation, IQ imbalance, DC offset) imperfections with varied severity levels.
  \item \textbf{Received SNR:} Wide distribution of ground-truth SNRs resulting from combined effects, captured in metadata.
  \item \textbf{Spectral Coexistence:} Scenarios ranging from sparse occupancy to dense environments with controlled signal overlap (up to 15\% probability/extent in reference config) simulating interference.
\end{itemize}
Table~\ref{tab:dataset_params_summary_revised} summarizes the parameter ranges used in the CSRD2025 generation based on the default configuration. This multi-faceted diversity makes CSRD2025 a rich resource for training robust and generalizable AI models.

\subsection{Data Formats and Primary Metadata}
\label{subsec:dataset_structure_format}

CSRD2025 organizes data systematically. Complex IQ signals are stored in `.mat' files within \path{<SaveFolder>/sequence_data/iq/}. Each IQ file has a corresponding JSON metadata file in \path{<SaveFolder>/anno/}, linked by filename (e.g., `Frame\_XXXXXX\_Rx\_YYYY.json'). This JSON file provides exhaustive ground truth about the simulation parameters used to generate that specific IQ recording, structured to capture essential information analogous to SigMF~\cite{SigMFStandardRef}.

The core structure, illustrated by the snippet below, includes an `annotation' object containing receiver (`rx') and transmitter (`tx') details:
\begin{lstlisting}[language=json, caption={Representative snippet of primary metadata structure.}, label={lst:json_metadata}]
{
  "annotation": {
    "rx": {
      "MasterClockRate": 1.11E+6,
      "NumReceiveAntennas": 3,
      "TimeDuration": 0.331...,
      "IqImbalanceConfig": {"A": 3.10..., "P": 3.61...},
      "MemoryLessNonlinearityConfig": {"Method": "Ghorbani model", ...},
      "ThermalNoiseConfig": {"NoiseTemperature": 180.34...},
      "SiteConfig": {"Name": "Rx_...", "Antenna": {"Height": 8, ...}},
      "SNRs": [ [1.98, 8.59], 7.79, ... ] % SNR per signal/link (nested for MIMO/SIMO)
      % ... other rx parameters ...
    },
    "tx": [ % Array of transmitters contributing to this rx frame
      {
        "ModulatorType": "GMSK",
        "ModulatorOrder": 2,
        "CarrierFrequency": 27000,
        "NumTransmitAntennas": 1,
        "MemoryLessNonlinearityConfig": {"Method": "Hyperbolic tangent", ...},
        "PhaseNoiseConfig": {"Level": -129, ...},
        "SiteConfig": {"Name": "Tx_...", "Antenna": {"Height": 16, ...}},
        "FadingDistribution": "Rayleigh",
        "MaximumDopplerShift": 0.0022...,
        "PathDelays": [0, 1E-7, ...],
        "AveragePathGains": [0, -3.33, ...],
        "StartTimes": [0.017, 0.096...],
        "TimeDurations": [0.013..., 0.013...],
        "BandWidth": [ [-16249, 16249], [-16340, 16340] ],
        % ... other tx signal/link parameters ...
      },
      {
        "ModulatorType": "SSBAM",
        "CarrierFrequency": 50000,
        "IsDigital": false,
        "FadingDistribution": "Rician",
        "KFactor": 2.58...,
        % ... other tx signal/link parameters ...
      },
      % ... potentially more transmitter entries ...
    ]
  },
  "filePrefix": "Frame_000083_Rx_0001"
}
\end{lstlisting}

As shown in Listing~\ref{lst:json_metadata}, the `rx' object details the receiver setup, including clock rates, antenna configuration, applied RF impairments, site configuration, and crucially, the calculated ground-truth `SNRs' for signals received. The `tx' array contains an entry for each transmitter contributing signals to this receiver frame. Each entry specifies the signal modulation, carrier frequency, transmitter impairments , site details, the specific channel parameters for the link to this receiver, and precise timing/bandwidth information. This detailed, structured metadata enables direct use of the IQ data for supervised learning and allows for fine-grained performance analysis based on specific generation parameters.

\begin{figure}[t]
  \centering
  \includegraphics[width=0.98\linewidth]{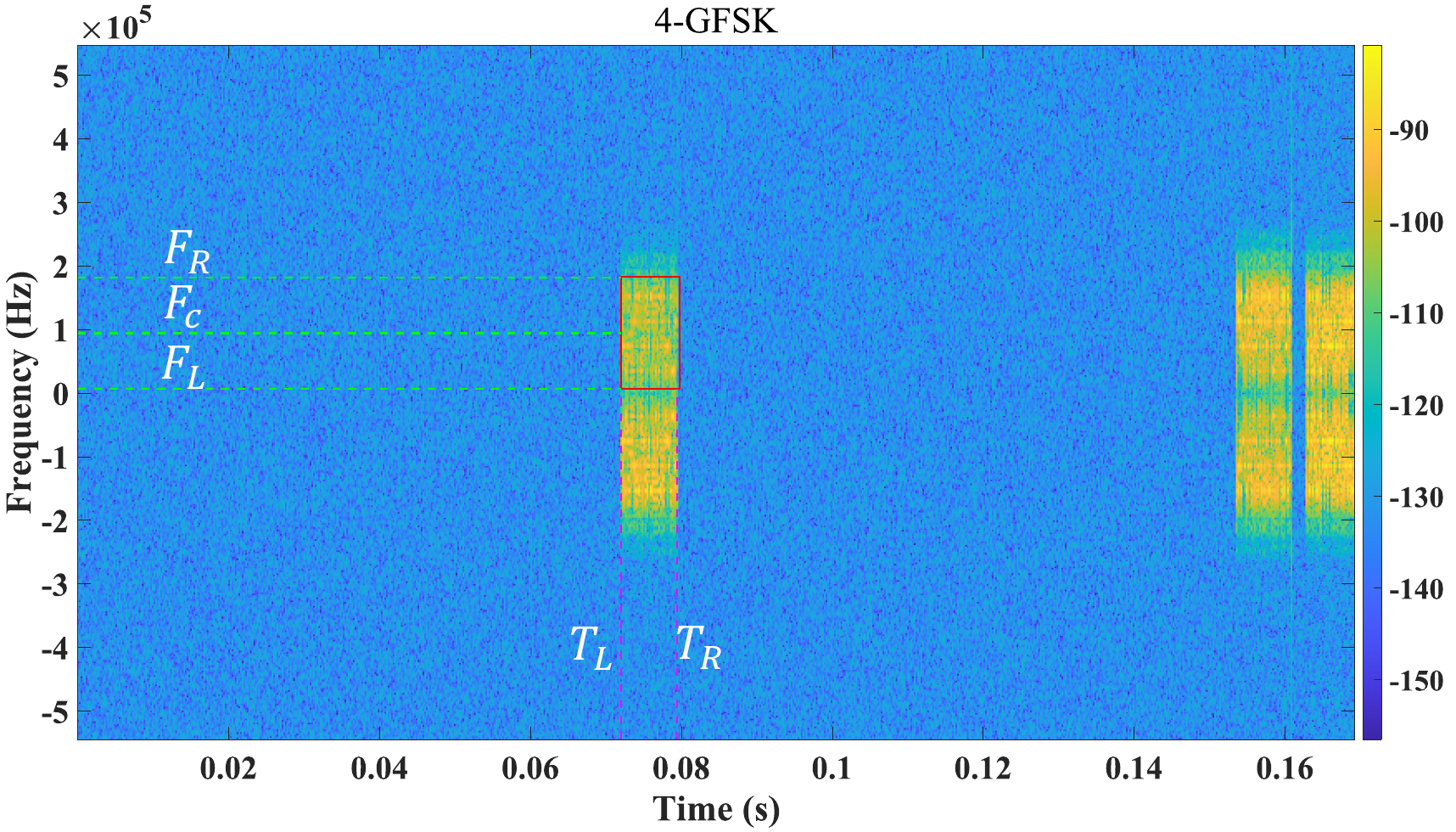}
  \caption{Example spectrogram generated from CSRD2025 data using STFT (Hamming window). A 4-GFSK signal instance is highlighted with a bounding box, indicating its time extent (from $T_L$ to $T_R$) and frequency range (approximated between $F_L$ and $F_R$, centered around $F_C$). This illustrates the representation used for object detection annotations.}
  \label{fig:spectrogram_example}
\end{figure}

To facilitate the application of object detection techniques for signal analysis in the time-frequency domain, we provide processing scripts and resulting annotations based on spectrogram representations of the IQ data. The complex IQ data from each `.mat' file is transformed into a spectrogram using the Short-Time Fourier Transform (STFT); the reference implementation utilizes a Hamming window or other configurable windowing functions. Figure~\ref{fig:spectrogram_example} shows an example spectrogram derived from a CSRD2025 IQ recording, representing signal power (or magnitude) across time bins (horizontal axis) and frequency bins (vertical axis).

\subsection{Spectrogram Representation and COCO Annotations for Object Detection}
\label{subsec:dataset_cocometa}

For object detection tasks, the ground truth information from the primary metadata is then translated into the widely used COCO (Common Objects in Context) annotation format~\cite{Lin2014MicrosoftCOCO}. As illustrated conceptually in Fig.~\ref{fig:spectrogram_example}, the time extent ($T_L$ to $T_R$) and frequency range ($F_L$ to $F_R$) of each signal are mapped to a bounding box defined by its top-left corner coordinates (`x', `y') and its dimensions (`width', `height') in the spectrogram image space. This conversion process results in standard COCO-style JSON files for each dataset split, detailing these bounding boxes and the corresponding class labels (modulation types, such as 4-GFSK as shown in the figure example) for all signal instances within the spectrograms. Readers are referred to the original COCO documentation for specifics of the format structure.

This COCO annotation structure makes CSRD2025 directly usable with standard object detection frameworks (e.g., Detectron2, MMDetection) for tasks like identifying the time-frequency location, duration, bandwidth, and modulation type of signals within a wideband spectrogram.

\subsection{Dataset Statistics}
\label{subsec:dataset_statistics}

Understanding the internal distribution of key parameters within CSRD2025 is crucial for interpreting model performance. Based on the reference generation configuration and the resulting primary and COCO metadata, we characterize the dataset as follows:

\begin{figure}[t]
  \centering
  \includegraphics[width=0.9\linewidth]{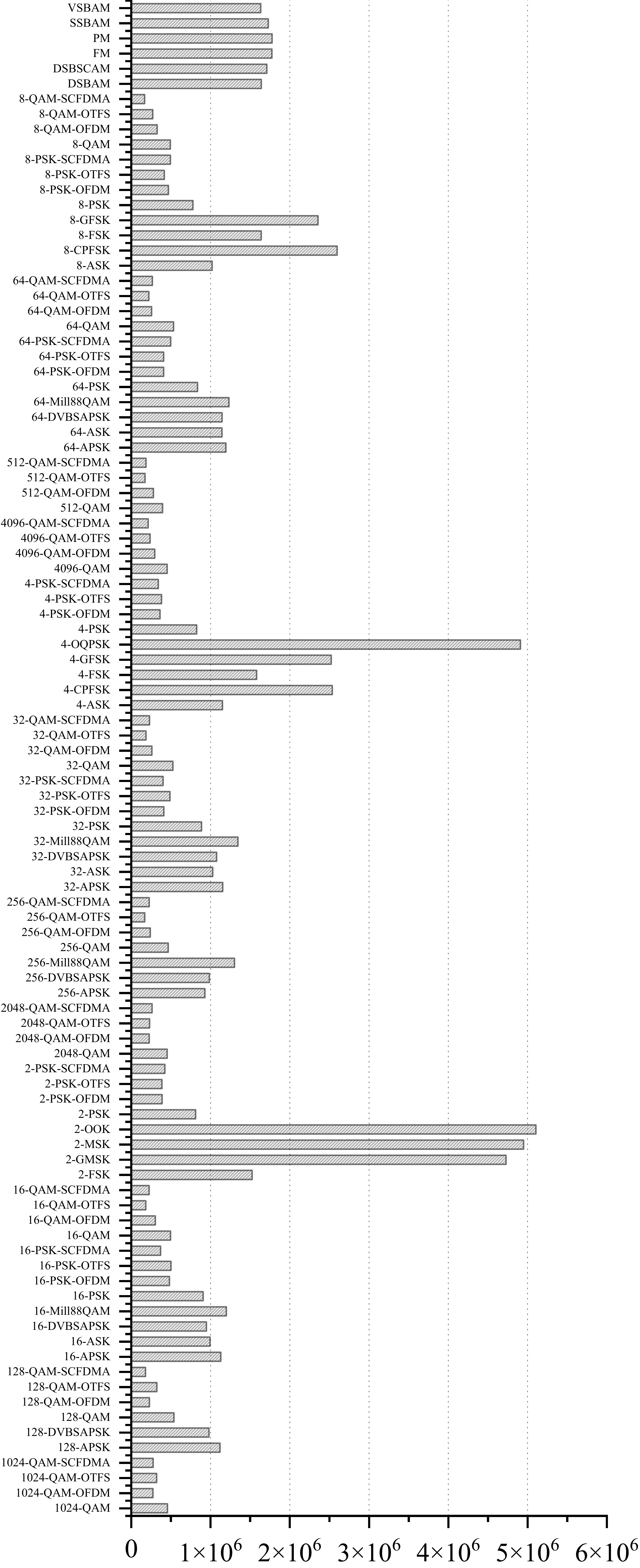}
  \caption{Distribution of signal instances per modulation class in the CSRD2025 training set (derived from COCO annotations).}
  \label{fig:mod_distribution_placeholder}
\end{figure}

\textbf{1) Modulation Class Distribution:} The CSRD2025 dataset encompasses \textbf{100 distinct modulation classes}, , offering a broad spectrum of analog, digital single-carrier, multi-carrier, and advanced multi-domain waveforms. For digital modulation schemes, the preceding numeral typically indicates the modulation order (e.g., '4-PSK' denotes Quadrature Phase Shift Keying). Figure~\ref{fig:mod_distribution_placeholder} presents the instance count per modulation class within the training split. The distribution reveals that foundational schemes like 2-OOK, 2-MSK, and 2-GMSK are most frequent (often $>4.5\times 10^6$ instances each). Many common digital and analog modulations (e.g., various PSK, FSK, CPM, QAM, AM, FM, PM) are also well-represented, typically with $1.5\times 10^6$ to $2.8\times 10^6$ instances. Conversely, more complex or specialized signals, such as very high-order QAMs (e.g., 1024-QAM and above) and many of their OTFS/SCFDMA variants, appear less frequently, generally having $2\times10^5$ to $5\times10^5$ instances. This tiered distribution ensures ample data for common signals while still providing diverse examples of more advanced and specialized waveforms, supporting the development of broadly capable LAMs.

\begin{figure}[t]
  \centering
  \includegraphics[width=0.9\linewidth]{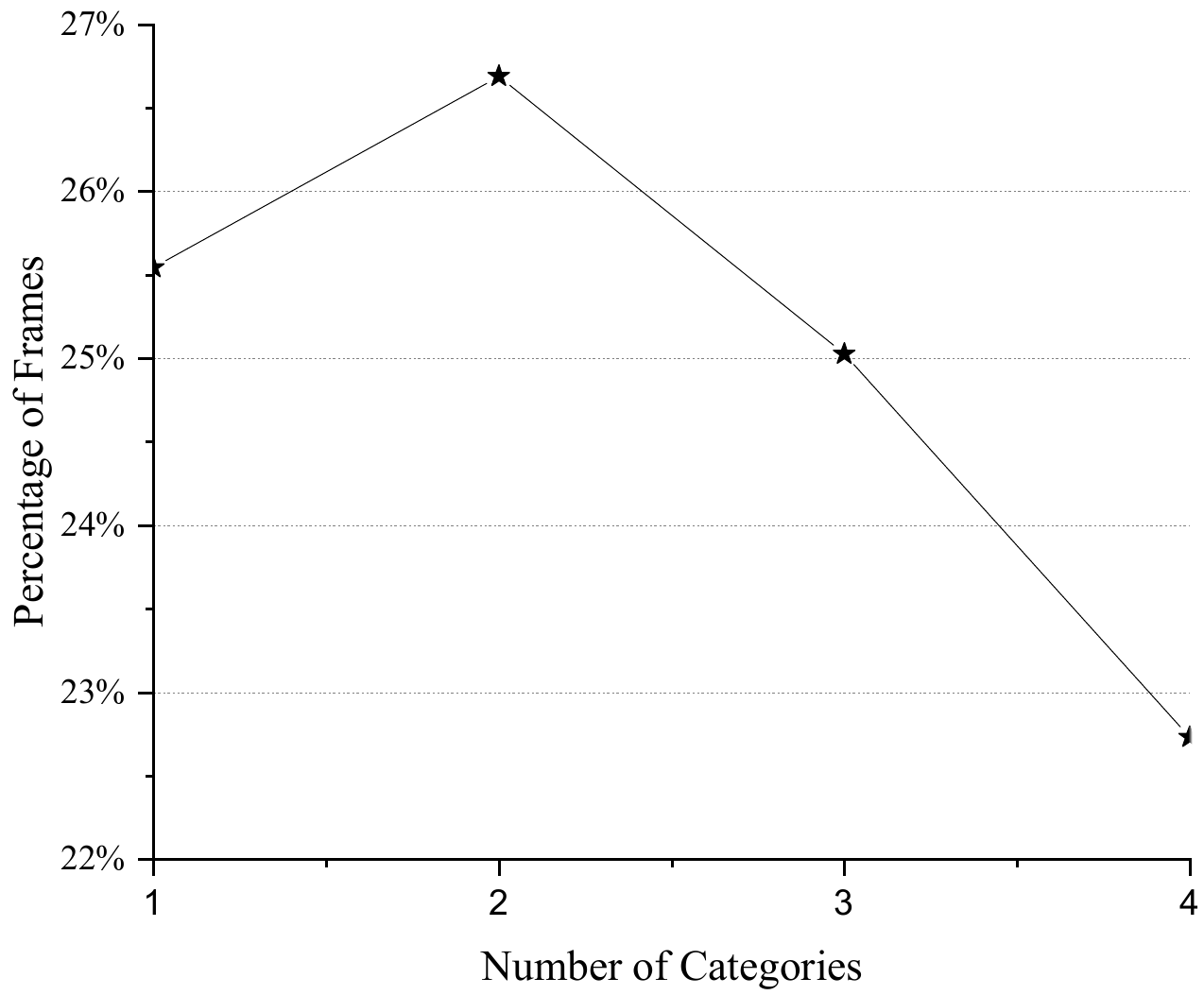}
  \caption{Distribution of the number of unique modulation categories per frame in the CSRD2025 training set.}
  \label{fig:categories_per_frame}
\end{figure}

\textbf{2) Signal Diversity per Frame:} To further characterize the complexity of the spectral scenes represented in CSRD2025, we analyzed the number of unique modulation categories and total signal instances present within each frame (COCO 'image') of the training set. Figure~\ref{fig:categories_per_frame} illustrates the distribution of the number of unique categories per frame. The data reveals that a vast majority of frames contain a mix of signal types: approximately 25.5\% of frames feature a single modulation category, 26.7\% contain two unique categories, 25.0\% have three, and 22.7\% present four distinct categories. This indicates a rich co-existence of different signal types within individual captured frames, providing diverse scenarios for training models to distinguish between multiple concurrent emissions.

\begin{figure}[t]
  \centering
  \includegraphics[width=0.9\linewidth]{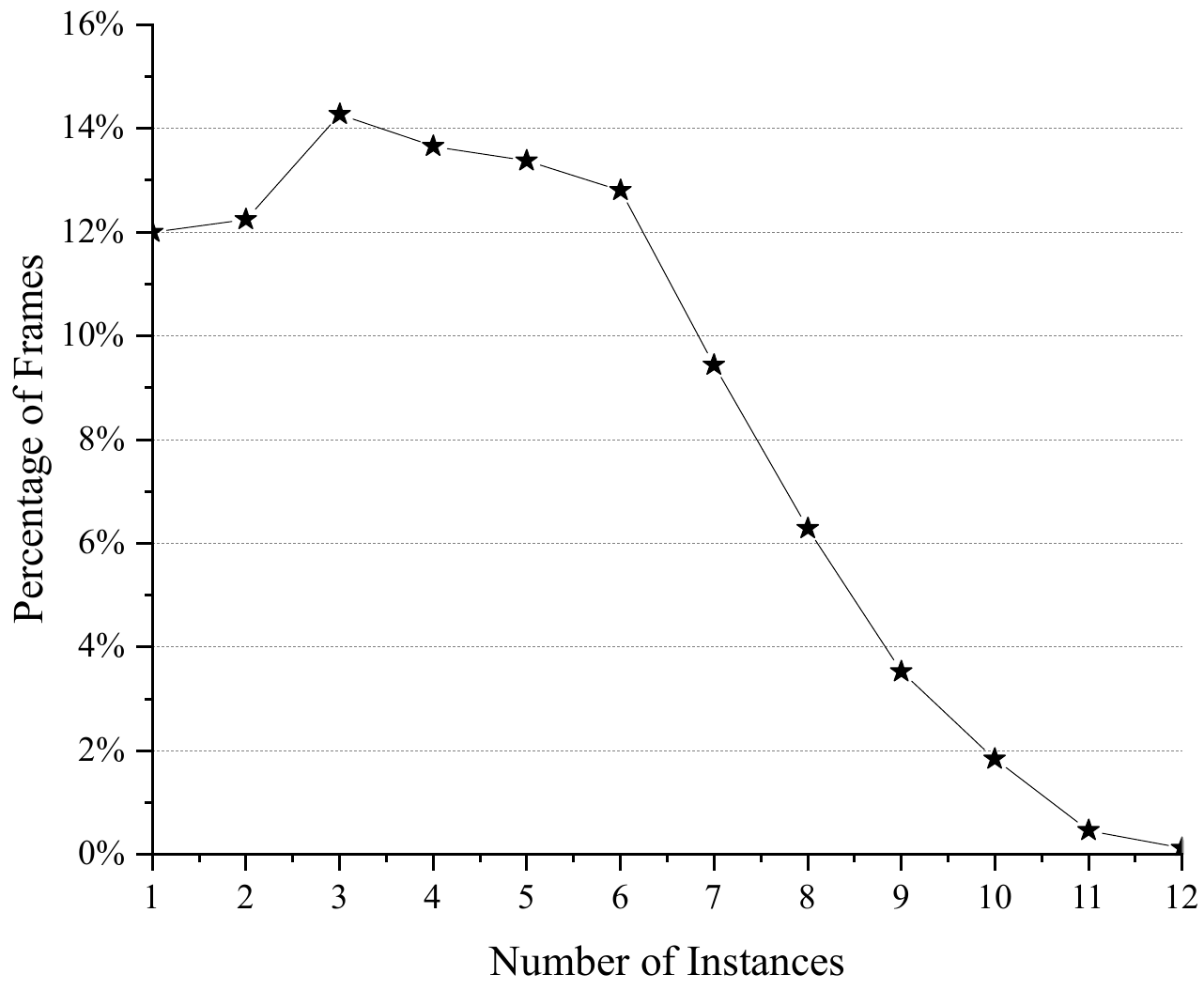}
  \caption{Distribution of the total number of signal instances per frame in the CSRD2025 training set.}
  \label{fig:instances_per_frame}
\end{figure}

\textbf{3) Signal Instance Density per Frame:} Beyond category diversity, the density of signal instances (i.e., the total number of individual signals, regardless of their modulation type) per frame was also examined, as shown in Figure~\ref{fig:instances_per_frame}. The distribution of instances per frame indicates that scenarios with a moderate number of signals are common. For example, frames containing 1, 2, 3, 4, 5, or 6 instances account for approximately 12.0\%, 12.2\%, 14.3\%, 13.7\%, 13.4\%, and 12.8\% of the training set, respectively. The proportion of frames decreases for higher instance counts, with frames containing 7, 8, 9, or 10 instances representing about 9.4\%, 6.3\%, 3.5\%, and 1.8\% respectively. Frames with 11 or 12 instances are less common (0.46\% and 0.12\%), and instances beyond this are even rarer. This distribution reflects the varying levels of spectral occupancy, from sparse to moderately dense, modeled within the dataset, which is critical for developing algorithms capable of handling different degrees of signal superposition and interference.

\begin{figure}[t]
  \centering
  \includegraphics[width=0.99\linewidth]{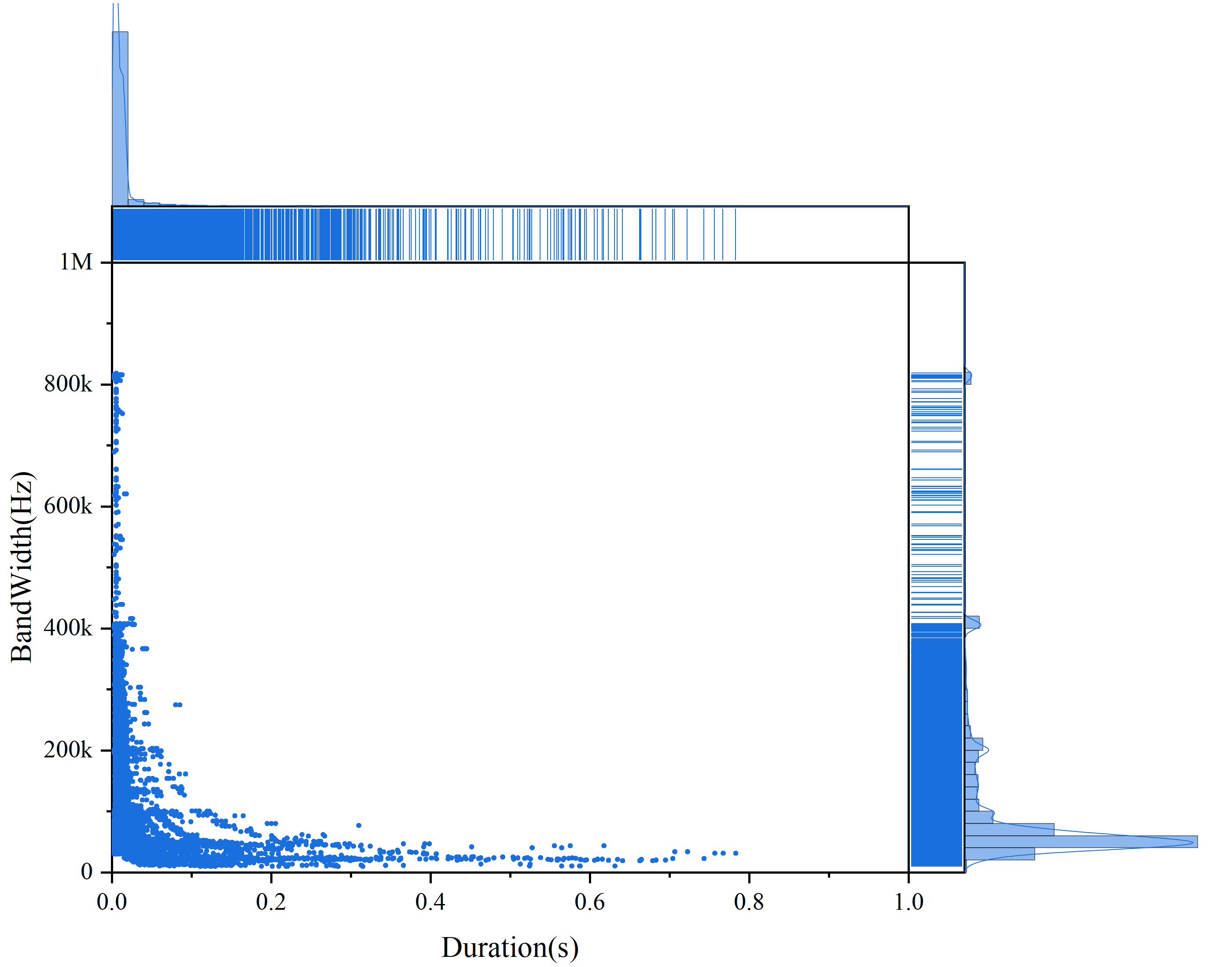}
  \caption{Distribution of signal duration and bandwidth in the CSRD2025 training set.}
  \label{fig:duration_bw_distribution}
\end{figure}

\textbf{4) Signal Duration and Bandwidth Distribution:}
The signals in CSRD2025 exhibit significant variation in both their time duration (bounding box width in spectrogram bins) and frequency occupancy (bounding box height, corresponding to actual signal bandwidth). This diversity is crucial for training robust object detection models. Figure~\ref{fig:duration_bw_distribution} aims to illustrate these distributions for the training set.

The distribution of signal durations (derived from primary metadata and exemplified by the provided frequency counts per duration bin, with durations typically in seconds) is heavily skewed towards shorter signals. In a representative sample, over 92\% of signal instances have a very short duration, falling within the 0.01-0.02 second bin. Instance counts decrease sharply for longer durations; for example, signals between 0.03-0.04 seconds constitute about 3.4\% of the instances, and this trend continues. While much rarer, the dataset does include signals with longer durations, extending up to approximately 0.8 seconds or more, forming a long tail in the distribution.

Similarly, the distribution of signal bandwidths (also derived from primary metadata) shows considerable variation, ranging from tens of kHz to over 800 kHz. A dominant characteristic is a very high concentration of signals (over 42\% of instances in the sample data) within the 50-60 kHz bandwidth range. Another significant peak (approx. 16\%) occurs in the 70-80 kHz range. While instance counts generally decrease for wider bandwidths, there are notable secondary peaks, such as around 410-420 kHz (approx. 2.7\% of instances) and 810-820 kHz (approx. 1.2\% of instances). The inclusion of signals with both common, moderate bandwidths and more specialized, wider bandwidths, combined with the diverse durations, ensures models are trained on a realistic representation of spectral occupancy. This variability in both time duration and frequency occupancy is essential for training object detection models that can robustly identify signals of different sizes and aspect ratios in the time-frequency domain.

\begin{figure}[t]
  \centering
  \includegraphics[width=0.9\linewidth]{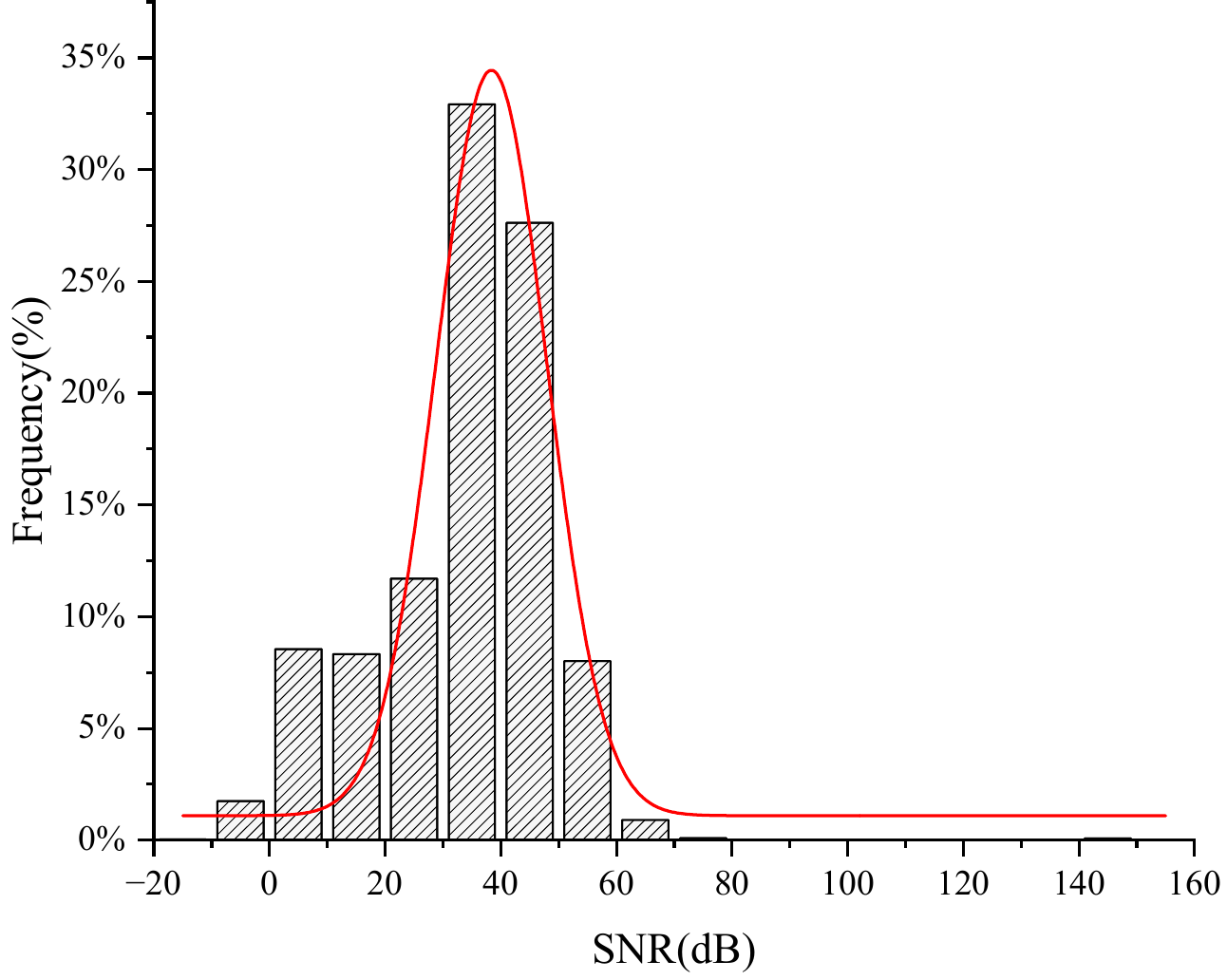}
  \caption{Distribution of per-signal ground-truth SNR across the CSRD2025 training set (derived from COCO annotations).}
  \label{fig:snr_distribution_placeholder}
\end{figure}

\textbf{5) SNR Distribution:} The ground-truth Signal-to-Noise Ratio (SNR) per signal, derived from the COCO annotations, spans a wide dynamic range designed to emulate diverse real-world reception conditions. Figure~\ref{fig:snr_distribution_placeholder} illustrates this distribution for the training split. The SNR values range from challenging low conditions (e.g., -15 dB and below) to very high SNR scenarios (extending beyond +70 dB, though instances become sparse at such extremes). The distribution is notably concentrated in the 0 dB to 50 dB range. Prominent peaks in instance counts are observed around +5 dB, +15 dB, and +25 dB, with the largest concentration of signals (nearly $3\times 10^7$ instances in the provided sample data) occurring in the +35 dB to +40 dB bin. This broad and varied SNR profile ensures that models trained on CSRD2025 are exposed to a realistic spectrum of signal strengths, crucial for developing robust AI-driven spectrum sensing algorithms.

These statistical characterizations provide essential context for utilizing the dataset and evaluating model performance across different conditions and signal types.

\subsection{Standardized Dataset Splits and Access}
\label{subsec:dataset_splits}

To facilitate consistent evaluation and reproducible research, CSRD2025 is released with pre-defined, fixed partitions for training, validation, and testing using a standard 8:1:1 ratio. 80\% of the simulation frames are allocated to the training set, 10\% to the validation set (for hyperparameter tuning), and 10\% to the test set (for final evaluation). This prevents discrepancies arising from different data splitting methods across studies. The specific frame indices for each split are provided alongside the dataset and framework release.

\section{Conclusion}
\label{sec:conclusion}
This paper addressed the critical need for large-scale, high-fidelity data for training LAMs in wireless spectrum sensing. We introduced the CSRD framework, an open-source, modular, and scalable simulation platform designed to generate realistic synthetic RF data. Using this framework, we created the CSRD2025 dataset. By providing these resources, we aim to significantly lower the barrier for entry into AI-driven wireless communications research. The CSRD framework and CSRD2025 dataset offer essential tools to accelerate the development, training, and evaluation of sophisticated AI models, ultimately contributing to the realization of more intelligent and adaptive spectrum sensing and management techniques for future wireless networks.

\section{Acknowledgement}
Thanks my wife Hui Li and my dear daughter Qing-Yue Chang for their love and support.
\bibliographystyle{IEEEtran}
\bibliography{ref.bib}



%\vfill


\end{document}